\documentclass[journal]{IEEEtran}

\usepackage{bm}
\usepackage[cmex10]{amsmath}
\usepackage{amssymb}
\usepackage{amsthm}
\usepackage{wasysym}

\usepackage{graphicx}

\usepackage{cite}

\theoremstyle{remark}
\newtheorem{remark}{Remark}

\theoremstyle{plain}

\usepackage{algorithm}
\usepackage{algorithmicx}
\usepackage{algpseudocode}

\graphicspath{{./}{./figs/}}

\begin{document}

\title{Distributed Variational Bayesian Algorithms for Extended Object Tracking}

\author{Junhao~Hua, Chunguang~Li,~\IEEEmembership{Senior~Member,~IEEE},
	\thanks{
	 The authors are with the Machine Intelligence Technology Lab, Alibaba Group; the College of Information Science and Electronic Engineering,  Zhejiang University, Hangzhou, China.}
	\thanks{E-mail: junhao.hjh@alibaba-inc.com, cgli@zju.edu.cn.}
}

	

\maketitle

\begin{abstract}		
This paper is concerned with the problem of distributed extended object tracking,
which aims to collaboratively estimate the state and extension of an object by a network of nodes.
In traditional tracking applications, most approaches consider an object as a point
source of measurements due to limited sensor resolution capabilities. 
Recently, some studies consider the extended objects, 
which are spatially structured, i.e., multiple resolution cells are occupied by an object.
In this setting, multiple measurements are generated by each object per time step.
In this paper, we present a Bayesian model for extended object tracking problem in a sensor network.
In this model, the object extension is represented by a symmetric positive definite random matrix,
and we assume that the measurement noise exists but is unknown.
Using this Bayesian model, we first propose a novel centralized algorithm for extended object tracking 
based on variational Bayesian methods. Then, we extend it to the distributed scenario 
based on the alternating direction method of multipliers (ADMM) technique.
The proposed algorithms can simultaneously estimate the extended object state (the kinematic state and extension)
and the measurement noise covariance. Simulations on both extended object tracking and group target tracking 
are given to verify the effectiveness of the proposed model and algorithms.
\end{abstract}
 

\IEEEpeerreviewmaketitle
  
\section{Introduction}

\IEEEPARstart{T}{racking} a moving target is crucial for many applications 
such as robotics, surveillance, monitoring, and security \cite{bar2004estimation, oh2007tracking,ribeiro2010kalman,li2016weighted,zhu2013distributed}.
In such scenarios, a sensor network can be deployed in order to increase the size of the surveillance area and cooperatively track targets.
Sensor networks consist of an amount of spatially distributed nodes that have limited communication capabilities due to energy and bandwidth constraints.
In recent years, the problem of distributed tracking over sensor networks, where nodes gather sensor data about one or multiple targets 
and then cooperatively estimate their current and future states using only local noisy measurements and information obtained from one-hop neighbors, has attracted a lot of attention 
 \cite{olfati2009kalman, cattivelli2010diffusion,chen2017weighted,hua2017distributed}.
Compared with the traditional centralized approach, distributed approach
does not need a powerful fusion center, so it is more flexible and provides robustness to
node and/or link failures.
Due to these merits, distributed tracking  has been used in a wide range of fields, including security and surveillance \cite{chong2003sensor,bokareva2006wireless},
environmental monitoring (tracking of weather patterns and pollutants) \cite{santini2008first},
and biology (tracking of populations or individual animals) \cite{mainwaring2002wireless}.

Distributed algorithms for cooperative tracking of target states in sensor networks using multiple sensor measurements 
have been extensively studied in recent literature. 
The most common approach for distributed tracking is using linear Kalman filter
based on average consensus \cite{olfati2005consensus, olfati2011collaborative, yu2009distributed},
and diffusion strategies \cite{cattivelli2010diffusion,hu2012diffusion}.
Besides, considering the case that the underlying states and/or sensor observation models 
are nonlinear and/or non-Gaussian, the distributed particle filters are developed based on the sequential Monte Carlo 
\cite{seifzadeh2015distributed,hlinka2013distributed, hlinka2012likelihood,dias2013cooperative, read2014distributed}.
Some authors also improve the algorithms by considering some realistic network.
In \cite{shi2010kalman},  Kalman filter  is designed over a packet-dropping network through a probabilistic approach.
In \cite{ zhu2013distributed}, a distributed optimal consensus filter is developed for heterogeneous sensor networks.

Previous studies on distributed tracking problems are focused on the scenario where
a target is considered as a point source of measurements and can only give rise to at most a single measurement per time step at each node.
They assume that the extension of an object is neglectable in comparison with sensor resolution and error.
However, since model applications require more and more detailed physical information about the objects for detection, tracking, classification and identification,
there is an increasing need for recognizing extended objects as individual units \cite{koch2008Bayesian,feldmann2011tracking}.
In this case, the traditional approach does not applicable as it has a significant loss of information.
With increased resolution capabilities of modern sensors, multiple measurements from an object can be obtained for each sensor/node,
and a better approach is to treat an object as an extended object (EO) with both the kinematic state (e.g., position, velocity, and acceleration) 
and physical extension (e.g., size, shape and orientation).
Also, a group of closely spaced targets can be considered as an extended object.
The aim of extended object tracking  (EOT) is to estimate both the kinematic state and extension.
In the literature, there are some studies on extended object tracking problem,
including multiple hypothesis tracking framework \cite{blackman1999design,koch1997multiple}
random finite sets approaches \cite{mahler2007statistical,mahler2014advances},
and random matrix framework \cite{koch2008Bayesian,feldmann2011tracking,orguner2012variational,granstrom2012phd,granstrom2014new,lan2016tracking}, etc.
However, to the best of our knowledge, the problem of distributed extended object tracking in a sensor network has not been studied yet.

In this paper, we consider the problem of distributed extended object tracking in sensor networks.
We formulate a distributed extended object model in a Bayesian framework.
The object extension is modelled as an ellipse represented by a positive definite matrix, called extension matrix \cite{koch2008Bayesian}.
The measurements gathered at each time scan are assumed to be distributed over the object with a Gaussian distribution whose covariance is related to the extension matrix.
Since in many cases statistical sensor error (noise) cannot be neglected when compared to object extension \cite{feldmann2011tracking}, 
we also consider the unknown measurement noise.
For tractability, we use latent variables to represent the underlying
noise-free measurements and build a complete measurement likelihood. 
Moreover, to obtain the recursive Bayesian filter,
we choose the conjugate priors of the extended object state (kinematic state vector and extension matrix) 
and noise covariance for this complete measurement likelihood.

Based on this model, 
we derive a novel distributed variational Bayesian algorithm for 
simultaneously estimating the extended object state and the measurement noise covariance in sensor networks.
In the prediction step, each node updates the predicted distribution of the extended object state locally.
In the measurement update step, each node infers the posteriors of the extended object state and the noise covariance 
in a collaborative manner.  In detail, the variational Bayesian (VB) methods \cite{beal2003variational,hua2016distributed} 
are applied to computing analytical approximations of these posteriors due to their intractability,
and  the alternating direction method of multipliers (ADMM) technique is applied to achieve distributed consensus.
To evaluate the effectiveness of the proposed algorithm,  
we empirically demonstrate the superior performance of 
the proposed algorithm on both extended object tracking and group target tracking scenarios.

The main contributions of this paper are summarized as follows.
\begin{itemize}
	\item We formulate a Bayesian model for distributed tracking of the extended object with both the kinematic state and extension in senor networks.
	We consider the case that the measurement noise exists and is unknown.
	\item We derive a novel centralized algorithm to simultaneously estimate the extended object state and measurement noise covariance
	based on variational Bayesian methods.	
	\item We propose a distributed algorithm for extended object tracking in sensor networks by integrating the ADMM technique into the VB iterative procedure.
\end{itemize}

The rest of the paper is organized as follows.
In Section \ref{sec_problem}, we formulate the problem of distributed Bayesian extended objects tracking,
and present the measurement model and dynamical model.
In Section \ref{sec_vbalg}, a variational Bayesian algorithm is derived
for iteratively estimating the extended object state and measurement noise covariance.
In Section \ref{sec_distributed}, a distributed
algorithm for extended object tracking is proposed based on the ADMM.
Numerical simulations are presented in Section \ref{sec_simulation}.
Finally, conclusion is drawn in Section \ref{sec_conclusion}.

\emph{Notations:} 
The superscript transposition $(\cdot)^T$ denotes transposition, and
$[\cdot]_{ij}([\cdot]_{i})$ denotes the $ij$-entry of a matrix ($i$-entry of a vector).
The operator $\mathbb{E}[\cdot]$ or $\langle\cdot \rangle$ denotes the expectation, $\operatorname{tr}(\cdot)$ denotes the trace operator,
$\mathcal{N}(\cdot)$ is the Gaussian distribution, $\mathcal{W}(\cdot)$ is the Wishart distribution,
and $\mathcal{IW}(\cdot)$ is the inverse Wishart distribution.
The definition of the Wishart and inverse Wishart distributions are given in Appendix \ref{wishart}. 
Other notations will be given if necessary.

\section{Network Model and  Bayesian Formulation of Extended Object Tracking}  \label{sec_problem} 
 
Let us consider a connected sensor network consisting of $N$ nodes
distributed over a geographic region. We use graphs to represent
networks. The considered undirected graph without a self-loop
$\mathcal{G} = (\mathcal{V}, \mathcal{E})$ consists of a set of
nodes $\mathcal{V} = \{1,2,...,N\}$ and a set of edges $\mathcal{E}$, 
where each edge $(k,l) \in \mathcal{E} $ connects
an unordered pair of distinct nodes. For each node $k \in \mathcal{V}$, let $\mathcal{N}_k = \{l|(k,l) \in \mathcal{E}, k \neq l\}$ 
be the set of its neighboring nodes (excluding node $k$ itself). 
At each time scan $t$, the node $k$ collects a set of $n_{k,t}$ measurements $Y_{k,t} = \{y_{k,t}^i\}_{i=1}^{n_{k,t}}$
with each individual measurement described by the vector $y_{k,t}^i \in \mathbb{R}^{d}$. 
Let $Y_t = \{Y_{k,t}\}_{k\in \mathcal{V}}$ denote measurements received from all nodes at time $t$.
The accumulated sensor data of node $k$ is denoted as $\mathcal{Y}_{k,t}=\{Y_{k,l} \}_{l=1}^t$,
and the accumulated sensor data from all nodes is denoted as $\mathcal{Y}_{t} = \{Y_l\}_{l=1}^t = \{\mathcal{Y}_{k,t}\}_{k\in\mathcal{V}}$.

Traditional target tracking approaches only consider a target as a point source of measurements.
In this paper, we suppose that a target has some object shape, which is described by a spatial model.
We use the random matrix model \cite{koch2008Bayesian} as the spatial model.
It models the kinematic target state by a random vector $x_{t}$ and the object extension
by a symmetric positive definite (SPD) random matrix $X_{t} \in \mathbb{R}^{d \times d}$, where $d$ is the dimension of the object.
The vector $x_{t}$ represents the position of an object and its motion properties, such as velocity, acceleration and turn-rate.
The SPD matrix $X_{t}$ implies that the object shape is approximated by an ellipse. 
The ellipse shape may seem limiting, however this model is applicable to many real scenarios,
such as pedestrian tracking using LIDAR \cite{granstrom2012phd} and tracking of boats and ships using marine radar \cite{granstrom2015gamma,lundgren2016variational}.

The aim of target tracking in a sensor network is to collaboratively estimate the extended object state using the measurements collected by a network of nodes.
In the following, we model the measurements and the state transition in a Bayesian methodology, and propose a variational Bayesian approach to 
estimate the extended object states in next section.

\subsection{Measurement Modelling}

The measurement model at node $k$ is assumed as 
\begin{equation} \label{measure_linear}
y_{k,t}^i = (H_t \otimes \mathbf{I}_d) x_{t} + \nu_{k,t}^i, i=1,\dots, n_k^t,
\end{equation}
where $\otimes$ stands for the Kronecker product, 
$H_t=[ 1\ 0 \ 0] \in \mathbb{R}^{1 \times 3} $ is the measurement matrix in one-dimensional space (assuming that only positions are measured and the state in one-dimensional space
is [position, velocity, acceleration]$^T$), $\mathbf{I}_d$ is an identity matrix,
and $\nu_{t}^i$ is a white Gaussian noise.
The matrix $(H_t \otimes \mathbf{I}_d)$ picks out the Cartesian position from the kinematic vector $x_t$.
As pointed out in \cite{feldmann2011tracking}, when the measurement noise is not negligible compared with the extension $X_t$, it is better
to consider both the measurement noise and the uncertainties in $X_t$ simultaneously. Therefore, we assume
\begin{equation} \label{measure_noise}
\nu_{k,t}^i \sim \mathcal{N}(0,s X_{t} + R),
\end{equation}
where $s$ is a real scalar describing the effect of $X_t$, and $R  \in \mathbb{R}^{d \times d} $ is an unknown measurement noise covariance.
Note that it is assumed that the covariance of measurement noise $R$ is unchanged with the time scan $t$.

From (\ref{measure_linear}) and (\ref{measure_noise}), it is concluded that the measurements $\{y_{k,t}^i\}$ are independent and identically Gaussian distributed 
(given the extended object state $x_{t}$, $X_{t}$, and measurement noise covariance $R$) as,
\begin{equation} \label{measure_pdf}
P(y_{k,t}^i|x_{t}, X_{t},R) = \mathcal{N} (y_{k,t}^i; (H_t \otimes \mathbf{I}_d) x_{t}, s X_{t} + R).
\end{equation}
The scaling factor $s$ can be used to describe different types of EOs. For example, 
as justified by  \cite{feldmann2011tracking}, $s=1/4$ indicates that
the scattering centers within an ellipse are uniformly distributed. 
If the point targets in the group are uniformly distributed spatially,
$s=1/4$ also can be used.

Due to the measurement noise covariance in (\ref{measure_pdf}), the exact analytical solution for the extended object state
can not be obtained using the measurement likelihood (\ref{measure_pdf}). Therefore,
following the approach in \cite{orguner2012variational}, we rewrite (\ref{measure_pdf}) as 
\begin{equation}
P(y_{k,t}^i|x_{t}, X_{t},R)  = \int P(y_{k,t}^i, z_{k,t}^i |x_{t}, X_{t},R )  d z_{k,t}^i,
\end{equation}
where the latent variable $z_{k,t}^i \in \mathbb{R}^{d}$  is introduced to represent an underlying noise-free measurement corresponding to the noisy measurement $y_{k,t}^i$.
Thus,  we have the complete measurement likelihood 
\begin{equation}\label{complete_lld}
\begin{split}
&P(y_{k,t}^i, z_{k,t}^i |x_{t}, X_{t},R ) = P(y_{k,t}^i|z_{k,t}^i,R) P(z_{k,t}^i|x_{t}, X_{t})\\
&= \mathcal{N} (y_{k,t}^i; z_{k,t}^i, R) \mathcal{N} ( z_{k,t}^i; (H_t \otimes \mathbf{I}_d) x_{t}, s X_{t}).
\end{split}
\end{equation}

For notation simplicity, let $Z_{k,t} = \{z_{k,t}^i\}_{i=1}^{n_{k,t}}$ denote the set of latent variables at node $k$, and 
$Z_t = \{ Z_{k,t}\}_{k \in \mathcal{V}}$ denote a set of latent variables from all nodes at time scan $t$.

\subsection{Object State Dynamics Modeling}

\begin{figure}[!t]
	\centering
	\includegraphics[width=2.8in]{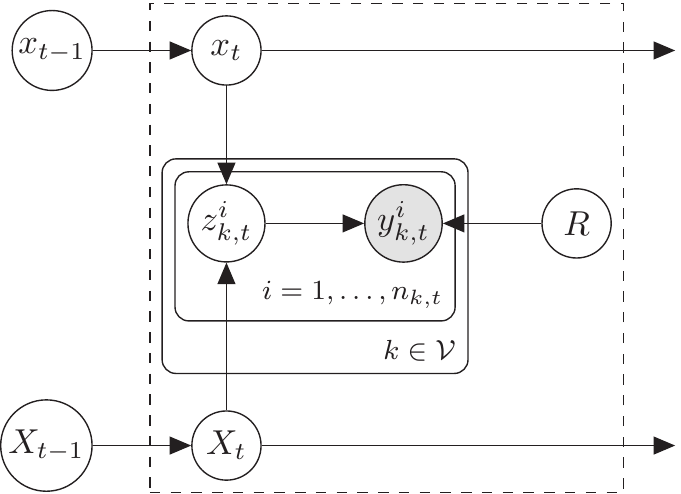}
	\caption{Graphical representation of the proposed distributed extended object model.}
	\label{fig_gm}
\end{figure}

The object dynamic model describes how the object state evolves over time.
In the context of the extended object tracking, this involves the descriptions 
of how the target kinematic state and the extension changed over time. 
We assume that the evolution of the extension is independent of the kinematical state,
and thus the transition density can be expressed as \cite{koch2008Bayesian},
\begin{equation}\label{evolu_model}
P(x_{t}, X_t | x_{t-1},X_{t-1}) = P(x_{t}|X_t, x_{t-1}) P(X_t | X_{t-1}).
\end{equation}
The object kinematic state has a linear Gaussian transition density 
\begin{equation} \label{kinmevolu}
P(x_{t}|X_t, x_{t-1}) = \mathcal{N}(x_{t}; (\mbox{F}_t \otimes \mathbf{I}_d) x_{t-1}, Q_t \otimes X_{t}),
\end{equation}
where $(\mbox{F}_t \otimes \mathbf{I}_d)  \in \mathbb{R}^{3d \times 3d}$ is the evolution matrix
with $\mbox{F}_t \in \mathbb{R}^{3 \times 3}$ being the dynamic matrix in the one-dimensional physical space, 
and $ (Q_t \otimes X_{t}) \in \mathbb{R}^{3d \times 3d}$ is the dynamics noise covariance.
We use the dynamic matrix $\mbox{F}_t$ given by \cite{koch2008Bayesian}
\begin{equation} 
\mbox{F}_t = 
\left(
\begin{array}{ccc}
1 & \Delta_t & \frac{1}{2} \Delta_t^2\\
0 & 1 & \Delta_t \\
0 & 0 & e^{-\Delta_t/\theta}
\end{array} \right),
\end{equation}
and the matrix $Q_t$ is given by van Keuk's model \cite{blackman1993phased}
\begin{equation}
Q_t = \Sigma^2 (1 - e^{-2\Delta_t/\theta})
\left(
\begin{array}{ccc}
0 & 0 & 0 \\
0 & 0 & 0 \\
0 & 0 & 1
\end{array}\right),
\end{equation}
with the scan time  $\Delta_t$, the scalar acceleration rms value $\Sigma$ 
and the maneuver correlation time constant $\theta$. 

We adopt the same assumption as \cite{koch2008Bayesian} that
the extension evolution has a Wishart transition density
\begin{equation} \label{dynamic_extension}
P(X_t | X_{t-1}) = \mathcal{W}_d(X_t; \eta_{t}, X_{t-1}/\eta_{t}),
\end{equation}
where the parameter $\eta_t>0$ governs the noise level of the prediction: the smaller $\eta_{t}$ is, the higher the process noise.


\subsection{Prior of the Extended object State} \label{sec_prior_state}
In the recursive Bayesian filtering, the prior of the extended object state is the predicted distribution $P(x_t,X_t | \mathcal{Y}_{t-1})$, 
and we typically want the prior and the posterior to be the same functional form. 
To ensure this, we should choose the conjugate prior for the complete measurement likelihood (\ref{complete_lld}).
Since the measurement likelihood (\ref{complete_lld}) is a product of two Gaussian distributions with unknown mean and unknown covariance, 
the conjugate prior of the extended object state is Gaussian inverse Wishart (GIW) distribution 
and the conjugate prior of the measurement noise covariance is inverse Wishart distribution.

Therefore, we first assume that the posterior of the extended object state at the time scan $t-1$ is Gaussian inverse Wishart (GIW) distributed,
 \begin{equation} \label{prev_update}
 \begin{split}
 P(x_{t-1},X_{t-1} | \mathcal{Y}_{t-1}) & =  P(x_{t-1} | X_{t-1}, \mathcal{Y}_{t-1}) P(X_{t-1} |  \mathcal{Y}_{t-1}) \\
 &=  \mathcal{N}(x_{t-1};m_{t-1|t-1}, P_{t-1|t-1} \otimes X_{t-1}) \\
 & \quad \times \mathcal{IW}_d(X_{t-1}; \nu_{k,t-1|t-1}, V_{t-1|t-1}),
 \end{split}
 \end{equation}
and then we set the predicted distribution $P(x_t,X_t | \mathcal{Y}_{t-1})$ as Gaussian inverse Wishart distributed presented in next section.
As for the measurement noise covariance $R$, we assume that it follows an inverse Wishart distribution at time scan $t-1$, i.e.,
\begin{equation} \label{prior_R}
P(R|\mathcal{Y}_{t-1}) = \mathcal{IW}(R; \upsilon_{t-1}, U_{t-1}).
\end{equation}

To capture the conditional dependence structure between random variables above, the graphical representation of the  
Bayesian extended object model is depicted in \figurename{\ref{fig_gm}}.

\section{Variational Bayesian Approach To Extended Object Tracking} \label{sec_vbalg}

For clarity, before presenting the distributed algorithm to estimate the extended object state in the considered network,
we would like to first present the corresponding centralized algorithm, in which at every time scan the measurements from all nodes
can be gathered in a fusion center and the computation is performed in the fusion center.
After that, we propose the distributed extended object tracking algorithm in next section.
Note that in the case when there is only a single node or measurements are collected at a central node, the centralized algorithm can be used.

The posterior of the extended object state at time scan $t$ can be recursively computed in two steps: the time update 
and the measurement update. In the time update step, we predict the distribution of the extended target state
based on the density $P(x_{t-1},X_{t-1} | \mathcal{Y}_{t-1})$ and the underlying evolution models (\ref{evolu_model}).
In the measurement update step, treating the predicted distribution as a prior,
we update the posterior of the extended target state based on the prior and the complete measurement likelihood (\ref{complete_lld}).

\subsection{Prediction}
The predicted distribution at time scan $t$ is defined as
\begin{equation} \label{pred_state}
\begin{split}
& P(x_{t},X_{t} | \mathcal{Y}_{t-1}) \\
& = \int P(x_{t}, X_t | x_{t-1},X_{t-1}) P(x_{t-1},X_{t-1} | \mathcal{Y}_{t-1}) d x_{t-1} d X_{t-1}.
\end{split}
\end{equation}
Using (\ref{evolu_model}) and (\ref{prev_update}),  the above predicted distribution can be factorized as
\begin{equation} \label{pred_prior_state}
\begin{split}
P(x_{t},X_{t} | \mathcal{Y}_{t-1}) = P(x_{t} | X_t, \mathcal{Y}_{t-1}) P(X_{t} | \mathcal{Y}_{t-1}),
\end{split}
\end{equation}
where
\begin{equation} \label{pred_x}
P(x_{t} | X_t, \mathcal{Y}_{t-1}) = \int P(x_{t}| X_t, x_{t-1}) P(x_{t-1} | X_{t}, \mathcal{Y}_{t-1}) d x_{t-1},
\end{equation}
and
\begin{equation} \label{pred_X}
P(X_{t} | \mathcal{Y}_{t-1}) = \int P(X_t | X_{t-1})  P(X_{t-1} | \mathcal{Y}_{t-1}) d X_{t-1}.
\end{equation}

\subsubsection{Kinematical State Part}
We adopt the assumption \cite{koch2008Bayesian} that 
$P(x_{t-1} | X_{t}, \mathcal{Y}_{t-1})$ follows a Gaussian distribution with the same parameters as $P(x_{t-1} | X_{t-1}, \mathcal{Y}_{t-1})$.
Since $ P(x_{t-1} | X_t, \mathcal{Y}_{t-1})$  and the transition density $P(x_{t}|X_t, x_{t-1})$ are both Gaussian, 
the predicted distribution $P(x_{t} | X_t, \mathcal{Y}_{t-1})$ in (\ref{pred_x}) is also Gaussian. Using the usual rules for Kronecker products,
we obtain
\begin{equation}  \label{pred_x_dist}
P(x_{t} | X_t, \mathcal{Y}_{t-1}) = \mathcal{N}(x_t; m_{t|t-1}, P_{t|t-1} \otimes X_t),
\end{equation}
where
\begin{equation}
\begin{split}
m_{t|t-1} &= (F_t \otimes \mathbf{I}_d) m_{t-1|t-1}, \\
P_{t|t-1} &= F_t P_{t-1|t-1} F_t^T + Q_t.
\end{split}
\end{equation}

\subsubsection{Object Extension Part}
Since $P(X_t | X_{t-1})$ is Wishart distributed and $P(X_{t-1} | \mathcal{Y}_{t-1})$ is inverse Wishart distributed, 
the predicted density $P(X_{t} | \mathcal{Y}_{t-1})$ in (\ref{pred_X}) is given by a ``Generalized Beta Type II'' density \cite{koch2008Bayesian}.
However, as we have mentioned in Section \ref{sec_prior_state}, we typically want
the predicted density (prior) $P(X_{t} | \mathcal{Y}_{t-1})$ to be the same functional form of its posterior.
Therefore, alternatively, we assume the predicted density $P(X_{t} | \mathcal{Y}_{t-1})$ is also inverse Wishart distributed,
and we use a heuristic approach to set its parameters.
In fact, according to \cite{koch2008Bayesian},  
if there is a sufficient number of sensor measurements,
the prediction part of the tracking process is unimportant compared with the gain obtained in the measurement update step.

Following \cite{feldmann2011tracking}, we postulate that the expectation of the extension is unchanged in the prediction, i.e., $\mathbb{E}[X_t] = \mathbb{E}[X_{t-1}]$,
and gradually decrease its degrees of freedom. Thus, the predicted density (\ref{pred_X})
is heuristically approximated by
\begin{equation} \label{pred_X_dist}
P(X_{t} | \mathcal{Y}_{t-1}) \approx  \mathcal{IW}(X_{t}; \nu_{t|t-1}, V_{t|t-1} ),
\end{equation}
where
\begin{subequations}
	\begin{align}
	\nu_{t|t-1} &= d  + 3 + e^{-\Delta_t /\tau} (\nu_{t-1|t-1} -d - 3), \label{nu_update} \\
	V_{t|t-1} &= \frac{\nu_{t|t-1}-d-1}{\nu_{t-1|t-1}-d-1} V_{t-1|t-1},
	\end{align}
\end{subequations}
with the temporal decay constant $\tau$ and scan time  $\Delta_t$ .

\subsection{Variational Bayesian Measurement Update}
In a fully Bayesian context, given the priors  (\ref{prior_R}) and (\ref{pred_prior_state}),
the aim of the measurement update is to compute the entire posterior distribution of the unobserved 
variables (extended object state $\{x_t, X_t\}$, the measurement noise covariance $R$ and the latent variables $Z_t$).
The posterior is given by Bayes' theorem,
\begin{equation}\label{bayes}
P(x_{t}, X_{t}, Z_{t},R |\mathcal{Y}_{t}) = \frac{P(Y_{t}, Z_{t}|x_{t}, X_{t}, R) P(x_{t}, X_{t}, R | \mathcal{Y}_{t-1})}{P(Y_t| \mathcal{Y}_{t-1})},
\end{equation}
where its denominator is the model evidence defined as
\begin{equation}
\begin{split}
& P(Y_t| \mathcal{Y}_{t-1}) \\
& = \int P(Y_{t}, Z_{t}|x_{t}, X_{t}, R) P(x_{t}, X_{t},R | \mathcal{Y}_{t-1}) d x_t d X_t d R dZ_t.
\end{split}
\end{equation}
However, the computation of exact posterior (\ref{bayes}) is intractable.
We apply an approximation technique, called variational Bayesian (VB) method \cite{beal2003variational}, for solving this problem.
The VB is to approximate the posterior $P(x_{t}, X_{t}, Z_t, R| \mathcal{Y}_{t})$ by a tractable distribution $Q(x_{t}, X_{t}, Z_t, R)$.
It is found by minimizing the Kullback-Leibler (KL) divergence between these two distributions \cite{beal2003variational},
\begin{equation}\label{kl}
\mbox{KL}(Q||P) = -\mathcal{L}(Q) + \ln P(Y_t|\mathcal{Y}_{t-1}),
\end{equation}
where $\mathcal{L}(Q)$ is the evidence lower bound of the marginal log-likelihood of the measurements $\ln P(Y_t|\mathcal{Y}_{t-1})$,
and it can be written as
\begin{equation}\label{lbd}
\mathcal{L}(Q) = \mathbb{E}_{Q(x_{t}, X_{t}, Z_t, R)} \left[\ln \frac{P(x_{t}, X_{t}, Z_t, R | \mathcal{Y}_{t})}{Q(x_{t}, X_{t}, Z_t, R)}\right].
\end{equation}

Since the log evidence $\ln P(Y_t|\mathcal{Y}_{t-1})$ is fixed with respect to $Q$, 
the minimization of (\ref{kl}) is equivalent to maximizing the lower bound (\ref{lbd}). 
To obtain an analytical approximate solution, we make the mean field assumption \cite{beal2003variational} and factorize
the joint variational distribution of $x_t$, $X_t$, $Z_t$ and $R$ as
\begin{equation}
Q(x_{t}, X_{t}, Z_t, R) = q(x_{t}, X_{t}) q(Z_t) q(R).
\end{equation}

Following \cite{hua2016distributed}, the global lower bound (\ref{lbd}) can be replaced by
an average of the local lower bounds, 
\begin{equation}\label{global_lower_bound}
\begin{split}
\mathcal{L}(Q) & = \mathbb{E}_q \left[\ln \frac{P(x_{t}, X_{t}, Z_t, R | \mathcal{Y}_{t})}
{q(x_t,X_t) q(R) \prod_{k=1}^{N} q(Z_{k,t})}\right] \\
& = \frac{1}{N}\sum_{k=1}^{N} \mathcal{L}_k(q),
\end{split}
\end{equation}
where $\mathcal{L}_k(q)$ is the local lower bound of each node $k$, defined as
\begin{equation} \label{localld}
\begin{split}
\mathcal{L}_k(q) \triangleq & \mathbb{E}_q \left[\ln \frac{ P(x_{t}, X_{t} | \mathcal{Y}_{t-1}) P(R |\mathcal{Y}_{t-1} ) } {q(x_t, X_t) q(R)}\right]\\
& + N \mathbb{E}_q \left[\ln \frac{P(Y_{k,t}|Z_{k,t}, R) P(Z_{k,t}|x_t, X_t)}{q(Z_{k,t})}\right].
\end{split}
\end{equation}
Note that this local objective function (\ref{localld}) only contains the local measurements $Y_{k,t}$ and latent variables $Z_{k,t}$,
and thus can be optimized locally at each node $k$.
To obtain ``best'' variational distribution of $Q$,
the VB alternates between maximizing the lower bound (\ref{global_lower_bound}) with respect to the 
variational distributions of the latent variables $\{Z_{k,t}\}$
and that of the parameters $x_t, X_t,R$, consisting of three iterative steps:
\begin{subequations}
	\begin{align}
	& q^*(Z_{k,t})  = \arg\max_{q_Z} \mathcal{L}_k(q^*(x_t,X_t), q(Z_{k,t}), q^*(R)), \label{vbm}	\\
	& q^*(x_t, X_t) = \arg\max_{q_{x_t,X_t}} \sum_{k=1}^{N} \mathcal{L}_k(q(x_t,X_t), q^*(Z_{k,t}), -), \label{vbe1}   \\	 	  	 	 
	& q^*(R)  = \arg\max_{q_R} \sum_{k=1}^{N} \mathcal{L}_k( -, q^*(Z_{k,t}), q(R)). \label{vbe2}
  	\end{align}
\end{subequations}

Based on the VB theory \cite{beal2003variational}, 
we can obtain the analytical solutions for variational distributions $q^*(Z_{k,t})$, $q^*(x_t, X_t)$ and $q^*(R)$.
In the following, we give these solutions respectively.

\subsubsection{Posterior of latent variables}

The optimal variational distribution $ q^*(z_{k,t}^i)$ has the form
\begin{equation}\label{qzform}
\begin{split}
& \ln q^*(z_{k,t}^i) = \mathbb{E}_{x_t, X_t, R}[\ln P(y_{k,t}^i, z_{k,t}^i |x_{t}, X_{t},R )] + c_z  \\
& =  \mathbb{E}_{x_t, X_t}[\ln  P(z_{k,t}^i|x_t, X_t)] +  \mathbb{E}_{R}[\ln P(y_{k,t}^i|z_{k,t}^i, R)] + c_z\\
& =  -\frac{1}{2} (z_{k,t}^i)^T\left( \langle (sX_{t})^{-1} \rangle + \langle R^{-1}  \rangle \right)z_{k,t}^i \\
& \quad + (z_{k,t}^i)^T \left( \langle R^{-1} \rangle y_{k,t}^i +  \langle (sX_{t})^{-1} \rangle (H_t \otimes \mathbf{I}_d) \langle x_t \rangle \right) + c_z, \\
\end{split}
\end{equation}
where $c_z$ is a constant term with respect to the variable $z_{k,t}^i$. 
From (\ref{qzform}), it is obvious that $ q^*(z_{k,t}^i)$ follows a Gaussian distribution,
\begin{equation} \label{q_zkti}
q^*(z_{k,t}^i) = \mathcal{N}(z_{k,t}^i; \hat{\mu}_{k,t}^i, \hat{\Sigma}_{k,t}^i),
\end{equation}
with the parameters given by
\begin{subequations}
	\begin{align}
	\hat{\mu}_{k,t}^i & = \hat{\Sigma}_{k,t}^i(\langle R^{-1} \rangle y_{k,t}^i + \frac{1}{s} \langle X_{t}^{-1}\rangle (H_t \otimes \mathbf{I}_d)  \langle x_t \rangle), \\
	\hat{\Sigma}_{k,t}^i & = (\langle R^{-1} \rangle + \frac{1}{s} \langle X_{t}^{-1}\rangle )^{-1}.
	\end{align}
\end{subequations}

The expected sufficient statistics for updating other variational distributions are as follows
\begin{subequations}
	\begin{align}
	& \langle z_{k,t}^i \rangle = \hat{\mu}_{k,t}^i, \\
	& \langle z_{k,t}^i (z_{k,t}^i)^T \rangle = \hat{\Sigma}_{k,t}^i + \hat{\mu}_{k,t}^i (\hat{\mu}_{k,t}^i)^T, \\
	& \langle (y_{k,t}^i-z_{k,t}^i)(y_{k,t}^i-z_{k,t}^i)^T \rangle \nonumber \\
	& = (y_{k,t}^i - \langle z_{k,t}^i \rangle)(y_{k,t}^i - \langle z_{k,t}^i \rangle)^T + \hat{\Sigma}_{k,t}^i.
	\end{align}
\end{subequations}

\subsubsection{Posterior of the extended object state}
The optimal variational distribution $q^*(x_t,X_t)$ can be expressed as 
\begin{equation} \label{qxXform}
\begin{split}
& \ln q^*(x_t, X_t)   =  \sum_{k=1}^N \mathbb{E}_{Z_{k,t}}[\ln P(Z_{k,t}|x_t, X_t) ] \\
				 & \quad + \ln P(x_{t} |X_t, \mathcal{Y}_{t-1}) + \ln P(X_t |  \mathcal{Y}_{t-1}) + c_{x}\\
				 & = - \frac{1}{2} \sum_{k=1}^N \sum_{i=1}^{n_{k,t}} \langle (z_{k,t}^i - (H_t \otimes \mathbf{I}_d)x_t)^T\\
				 & \qquad \qquad \qquad \times  (s X_t)^{-1} (z_{k,t}^i -  (H_t \otimes \mathbf{I}_d)x_t) \rangle \\
				 & \quad  -\frac{\sum_{k=1}^{N} n_{k,t}}{2} \ln |s X_t| -\frac{1}{2} \ln |P_{t|t-1} \otimes X_t| \\				 
				  & \quad  -\frac{1}{2}(x_t - m_{t|t-1})^T (P_{t|t-1} \otimes X_t)^{-1}(x_t - m_{t|t-1}) \\
				  &  \quad - \frac{\nu_{t|t-1} + d + 1}{2} \ln |X_t| - \frac{1}{2} \operatorname{tr}(X_t^{-1} V_{t|t-1}) + c_{x},\\		
\end{split}
\end{equation}
where $c_x$ denotes a constant term with respect to $x_t$ and $X_t$. 
In the Appendix \ref{appendix_proof_xX}, we proved that (\ref{qxXform}) can be rewritten as 
\begin{equation} \label{qxX}
\begin{split}
q^*(x_t, X_t) & = q^*(x_t|X_t)q^*(X_t) \\
& = \mathcal{N}(x_t; \hat{m}_t, \hat{P}_t \otimes X_t) \mathcal{IW}(X_t; \hat{\nu}_t, \hat{V}_t),
\end{split}
\end{equation}
where  the parameters are given by
\begin{subequations}\label{param_Xx}
	\begin{align}
		\hat{m}_t & =   m_{t|t-1} +   ( w_t \otimes \mathbf{I}_d )( \bar{z}_t -  (H_t \otimes \mathbf{I}_d) m_{t|t-1} ), \label{mt}\\
	\hat{P}_t  & = P_{t|t-1} -  b_t w_t  w_t^T, \label{hatpt} \\
	\hat{\nu}_t & = \nu_{t|t-1} + N_t, \label{hatnut}\\
	\hat{V}_t  & =  V_{t|t-1} + \frac{ N_t }{s} \mathbf{S}_t + \mathbf{K}_t. \label{Vt}
	\end{align}
\end{subequations}
In (\ref{param_Xx}), $b_t$ is a scalar innovation factor, $w_t$ is a gain vector defined as
\begin{subequations}
	\begin{align}
	b_t & \triangleq \frac{s}{ N_t } + H_t P_{t|t-1}H_t^T,\\
	w_t & \triangleq P_{t|t-1}H_t^T b_t^{-1}, 	
	\end{align}
\end{subequations}
and the corresponding statistics are given by
\begin{subequations}
	\begin{align}
	N_t  &  \triangleq \sum_{k=1}^N n_{k,t},   \quad 
	\bar{z}_t   \triangleq \frac{1}{ N_t } \sum_{k=1}^N \sum_{i=1}^{n_{k,t}} \langle z_{k,t}^i \rangle, \\
	\mathbf{S}_t & \triangleq \frac{1}{ N_t } \sum_{k=1}^N \sum_{i=1}^{n_{k,t}}  \langle z_{k,t}^i (z_{k,t}^i)^T \rangle - \bar{z}_t \bar{z}_t ^T,  \\	
	\mathbf{K}_t & \triangleq b_t^{-1}(\bar{z}_t - (H_t \otimes \mathbf{I}_d) m_{t|t-1})(\bar{z}_t - (H_t \otimes \mathbf{I}_d) m_{t|t-1})^T.
	\end{align}
\end{subequations}

The expected sufficient statistics of $q^*(x_t, X_t)$ are
\begin{subequations}
	\begin{align}
	& \langle x_t \rangle = \hat{m}_t, \\
	& \langle X_t \rangle = \frac{\hat{V}_t}{\hat{\nu}_t-d-1}, \langle X_t^{-1} \rangle  = \hat{\nu}_t \hat{V}_t^{-1}.
	\end{align}
\end{subequations}

\subsubsection{Posterior of the measurement noise covariance}
The optimal variational posterior $ q^*(R)$ has the form
\begin{equation}\label{qRform}
\begin{split}
& \ln q^*(R) \\
&  = \sum_{k=1}^N \mathbb{E}_{Z_{k,t}}[\ln P(Y_{k,t}|Z_{k,t}, R) ] + \ln  P(R | \mathcal{Y}_{t-1})  + c_{R}\\
		   & = - \frac{ (N_t + \upsilon_{t-1}) + d + 1}{2} \ln |R| \\
			 & - \frac{1}{2} \operatorname{tr}( R^{-1} (\sum_{k=1}^{N}\sum_{i=1}^{n_{k,t}} \langle (y_{k,t}^i-z_{k,t}^i)(y_{k,t}^i-z_{k,t}^i)^T\rangle  + U_t)) + c_{R},
\end{split}
\end{equation}
where $c_R$ denotes a constant term with respect to $R$.
From (\ref{qRform}), we conclude that $q^*(R)$ follows a inverse Wishart distribution,
\begin{equation} \label{q_R}
q^*(R) = \mathcal{IW}(R; \hat{\upsilon}_t, \hat{U}_t),
\end{equation}
with the parameters given by
\begin{subequations}
	\begin{align}
	\hat{\upsilon}_t & =  \upsilon_{t-1} + N_t, \\
	\hat{U}_t & =  U_{t-1} + \sum_{k=1}^{N}\sum_{i=1}^{n_{k,t}} \langle (y_{k,t}^i-z_{k,t}^i)(y_{k,t}^i-z_{k,t}^i)^T  \label{Ut}\rangle. 
	\end{align}
\end{subequations}
The expected sufficient statistics of $q^*(R)$ are 
\begin{equation}
\langle R \rangle =\frac{\hat{U}_t}{\hat{\upsilon}_{t}-d-1}, \langle R^{-1} \rangle = \hat{\upsilon}_{t} \hat{U}_t^{-1}.
\end{equation}

Using a superscript $(n)$ to denote the iteration number of the VB, starting from some initial parameters, the VB
alternates between the VBE step and VBM step. In the VBE step, the VB computes $q^{(n)}(Z_{k,t})$ using the expected sufficient statistics of $q^{(n-1)}(x_t, X_t)$ and $q^{(n-1)}(R)$. 
In the VBM step, the VB computes $q^{(n)}(x_t, X_t)$ and $q^{(n)}(R)$ using the expected sufficient statistics of $q^{(n)}(Z_{k,t})$.
We set the initial parameters as: 
$\bar{y}_{t} = \frac{1}{N_t}\sum_{k=1}^{N}\sum_{i=1}^{n_{k,t}}  y_{k,t}^{i}$, 
$m_{t}^{(0)} = (H_t^T \otimes \mathbf{I}_d) \bar{y}_{t}$, 
$P_{t}^{(0)} = P_{t|t-1}$,$V_{t}^{(0)} = 0.1\mathbf{I}_d$, ${\nu}_t^{(0)} =d+1+0.1$, 
$U_{t}^{(0)} = U_{t-1} $, ${\upsilon}_{t}^{(0)} = \upsilon_{t-1}$.

For clarity, the centralized variational Bayesian algorithm for extended object tracking (cVBEOT) is summarized in Algorithm \ref{alg_cEOT}.
Note that computing  a distribution means computing  its parameters.
The inner loop terminates when the VB algorithm converges or a predefined stopping criterion (e.g., a maximum iteration number) is satisfied.

\begin{algorithm}[t]
	\caption{Centralized Variational Bayesian Algorithm for Extended Object Tracking (cVBEOT) \label{alg_cEOT}}
	\begin{algorithmic}[0]		

		\Require{$P_{1|0}=\mathbf{I}_d$, $V_{1|0} = 0.1\mathbf{I}_d$, $\nu_{1|0} = d+1+0.1$,
			$U_{0} = 10^{-4}\mathbf{I}_d$, $\upsilon_{0} = d + 1$.}
		\Statex 
		\For{$t \gets 1,2,\dots$} \Comment{time scan}
		\State \emph{VB measurement update}:
		\State Initialization: $m_{t}^{(0)}, P_{t}^{(0)}, {V}_{t}^{(0)}, {\nu}_t^{(0)}, U_{t}^{(0)}, {\upsilon}_{t}^{(0)}$.
			\For{$ n \gets 1,2,\dots$}				
				\State Compute $\mathcal{N}(z_{k,t}^{i};\mu_{k,t}^{i,(n)}, \Sigma_{k,t}^{i,(n)}), \forall i, \forall k$ via (\ref{q_zkti}).
				\State  Compute $\mathcal{N}(x_{t}; m^{(n)}_{t}, P_{t}^{(n)} \otimes X_{t})$
				\State \quad \qquad $ \times \mathcal{IW}(X_{t}; \nu^{(n)}_{t}, V_{t}^{(n)})$ via (\ref{qxX}).
				\State Compute  $\mathcal{IW}(R; {\upsilon}_{t}^{(n)}, U_{t}^{(n)})$ via (\ref{q_R}).
			\EndFor		
			\State Update $m_{t|t} = m_{t}^{(n)}$, $P_{t|t} = P_{t}^{(n)}$.
			\State Update $\nu_{t|t} = \nu_{t}^{(n)}$, $V_{t|t} = V_{t}^{(n)}$.
			\State Update $\upsilon_{t|t} = \upsilon_{t}^{(n)}$, $U_{t|t} = U_{t}^{(n)}$.
			\State \emph{Prediction}: 
			\State	Compute  $P(x_{t+1},X_{t+1} | \mathcal{Y}_{t})$ via (\ref{pred_x_dist}) and (\ref{pred_X_dist}).
		\EndFor 				
	\end{algorithmic}
\end{algorithm}

\section{Distributed Variational Bayesian Algorithm for Extended Object Tracking} \label{sec_distributed}
In this section, we extend the centralized algorithm to the distributed  scenario 
where every node in the network only communicates with its one-hop neighboring nodes.
The main difficult in doing this is that the updating of the parameters $x_t$, $X_t$ and $R$ 
needs all expected sufficient statistics of latent variables $\{Z_{k,t}\}_{k \in \mathcal{V}}$.
To collect all these statistics, a possible solution is 
to find a cyclic path through all the nodes. However, it is not suitable for 
a low-cost networked system, since the communication resources are limited and exploring 
the network topology is hard and expensive.
To solve this problem, we present a fully distributed VB algorithm for the considered network.
Consider replacing the common extended object states $x_t$, $X_t$ and measurement noise covariances $R$ with a set of per node variables $\{x_{k,t}, X_{k,t}, R_{k}\}$,
we solve this problem by making an agreement on these parameters $\{x_{k,t}, X_{k,t}, R_{k}\}$ among all nodes.
The proposed algorithm consists of two steps: the local prediction and the distributed VB measurement update.
In the following, we present these two steps respectively.

\subsection{Local Prediction}
Since the dynamic matrices $F_t$ and $Q_t$ are the same among all nodes,
the prediction step  can be performed locally. 
The predicted distribution of extended object state at node $k$ is the same as its centralized counterpart in (\ref{pred_x_dist}) and (\ref{pred_X_dist}),
namely
\begin{equation} \label{local_pred}
\begin{split}
P(x_{k,t},X_{k,t} | \mathcal{Y}_{t-1}) & = \mathcal{N}(x_{k,t}; m_{k,t|t-1}, P_{k,t|t-1} \otimes X_{k,t}) \\
& \qquad \times \mathcal{IW}(X_{k,t}; \nu_{k,t|t-1}, V_{k,t|t-1} ),
\end{split}
\end{equation}
where the parameters are given by
\begin{subequations} \label{local_pred_params}
	\begin{align}
	m_{k,t|t-1} &= (F_t \otimes \mathbf{I}_d) m_{k,t-1|t-1}, \\
	P_{k,t|t-1} &= F_t P_{k,t-1|t-1} F_t^T + Q_t, \\
	\nu_{k,t|t-1} &= d  + 3 + e^{-\Delta_t/\tau} (\nu_{k,t-1|t-1} -d - 3), \\
	V_{k,t|t-1} &= \frac{\nu_{k,t|t-1}-d-1}{\nu_{k,t-1|t-1}-d-1} V_{k,t-1|t-1}.
	\end{align}
\end{subequations}

\subsection{Distributed Variational Bayesian Measurement Update}
After performing the local prediction step, each node collaboratively updates the kinematic state $x_{k,t}$, extension $X_{k,t}$ and measurement noise covariance $R_k$ using its 
local measurements $\{y_{k,t}^i\}$ collected at time scan $t$ and the information obtained from its neighboring nodes. 
The distributed variational Bayesian measurement update  consists of three iterative steps:
the VBE step, the consensus step, and the VBM step. 
In the following, we present these three steps respectively.

\subsubsection{VBE step}
In the VBE step, the posteriors of the local latent variables $\{z_{k,t}^i\}$ at node $k$ are computed using the expected sufficient statistics of the local extended object state $x_{k,t}, X_{k,t}$
and local noise covariance $R_k$. Following (\ref{q_zkti}), the variational distribution  $q^*(z_{k,t}^i)$, $\forall i=1,\dots, n_{k,t}$, is 
\begin{equation} \label{qzk}
q^*(z_{k,t}^i) = \mathcal{N}(z_{k,t}^i | \hat{\mu}_{k,t}^i, \hat{\Sigma}_{k,t}^i),
\end{equation}
with the parameters given by
\begin{subequations}\label{qzparams}
	\begin{align}
	\hat{\mu}_{k,t}^i & = \hat{\Sigma}_{k,t}^i( \hat{\upsilon}_{k,t} \hat{U}_{k,t}^{-1}  y_{k,t}^i 
	+ \frac{ \hat{\nu}_{k,t}}{s} \hat{V}_{k,t}^{-1} (H_t \otimes \mathbf{I}_d)  \hat{m}_{k,t}), \label{qzmu} \\
	\hat{\Sigma}_{k,t}^i & = (\hat{\upsilon}_{k,t} \hat{U}_{k,t}^{-1} + \frac{\hat{\nu}_{k,t}}{s} \hat{V}_{k,t}^{-1} )^{-1}. \label{qzsigma}
	\end{align}
\end{subequations}

\subsubsection{Consensus step}
We observe that the computation of variational distributions of the global variables $x_{k,t}, X_{k,t}, R_k$ 
only needs an average of all expected sufficient statistics of latent variables $\{Z_{k,t}\}$.
In detail, let us define a set of expected sufficient statistics of local latent variables as
\begin{equation} \label{omegakt}
\omega_{k,t} \triangleq [(\omega_{k,t}^1)^T, (\omega_{k,t}^2)^T, (\omega_{k,t}^3)^T ]^T, \forall k \in \mathcal{V},
\end{equation}
where 
\begin{equation}  
\begin{split}
& \omega_{k,t}^1 \triangleq  \frac{N}{N_t}\sum_{i=1}^{n_{k,t}} \langle z_{k,t}^i \rangle,  \ 
\omega_{k,t}^2 \triangleq  \frac{N}{N_t} \sum_{i=1}^{n_{k,t}} \langle z_{k,t}^i (z_{k,t}^i)^T \rangle,
\\
& \omega_{k,t}^3 \triangleq \sum_{i=1}^{n_{k,t}}   \langle (y_{k,t}^i-z_{k,t}^i)(y_{k,t}^i-z_{k,t}^i)^T \rangle.
\end{split}
\end{equation}
The computation of variational distributions $q^*(x_{k,t}, X_{k,t})$ and $q^*(R_k)$ then only needs an average of 
all local expected sufficient statistics, i.e.,
\begin{equation} \label{phi_avg}
\bar{\phi}_{t} \triangleq \frac{1}{N} \sum_{k=1}^N \omega_{k,t}.
\end{equation}

Our aim now becomes to compute (\ref{phi_avg}) in a distributed manner, which results in a distributed averaging problem \cite{olfati2005consensus,hua2016distributed}.
we use the ADMM technique \cite{boyd2011distributed} to solve this problem.
Let us define a set of per node intermediate quantities $\{\phi_{k,t}\}_{k=1}^N$ and 
add consensus constraints to force these variables to agree across neighboring nodes.
Thus, we could obtain a consensus-based optimization problem, 
\begin{equation}\label{cons_prob}
\begin{split}
&\min_{\{\phi_{k,t}\}, \{\varphi_{kj}\}} \frac{1}{2} \sum_{k=1}^N || \phi_{k,t} - \omega_{k,t}||_F^2, \\
& \mbox{s.t.} \quad \phi_{k,t} = \varphi_{kj}, \varphi_{kj} = \phi_{j,t}, \forall k \in \mathcal{V}, j \in \mathcal{N}_k,
\end{split}
\end{equation}
whose optimal value is equal to (\ref{phi_avg}).
In (\ref{cons_prob}), the auxiliary variable $\varphi_{kj}$ decouples local
variable $\phi_{k,t}$ at node $k$ from those of its neighbors $j \in \mathcal{N}_k$.
Let $\bm{\lambda}_{kj1}$ $(\bm{\lambda}_{jk2})$ denote the Lagrange multiplier corresponding 
to the constraint $\phi_{k,t} = \varphi_{kj}$ (respectively $\varphi_{kj} = \phi_{j,t}$),
and we construct the augmented Lagrangian function for the problem (\ref{cons_prob}) as follows,
\begin{equation}
\begin{split}
&\mathcal{L}_{\rho}(\{\phi_{k,t}\}, \{\varphi_{kj}\}, \{\bm{\lambda}_{kj\cdot}\}) = \\
& \frac{1}{2} \sum_{k=1}^{N} \Big( || \phi_{k,t} - \omega_{k,t}||_F^2 
+ \rho \sum_{j \in \mathcal{N}_k} || \phi_{k,t} - \varphi_{kj} + \bm{\lambda}_{kj1}||_F^2 \Big. \\
& \Big. \quad + \rho \sum_{j \in \mathcal{N}_k} ||  \varphi_{kj} - \phi_{j,t} + \bm{\lambda}_{kj2}||_F^2 \Big),
\end{split}
\end{equation}
where $\rho > 0$ is a penalty parameter. The ADMM cyclically minimizes $\mathcal{L}_{\rho}$
with respect to the local variables $\{\phi_{k,t}\}$ and auxiliary variables $\{\varphi_{kj}\}$,
followed by a gradient ascent step over the dual variables $\{\bm{\lambda}_{kj1}, \bm{\lambda}_{kj2}\}$.
Initializing all the Lagrange multipliers to zeros, the auxiliary variables $\{\varphi_{kj}\}$ can be expressed by $\{\phi_{k,t}\}$,
and we can obtain the iterations
required by per node $k$ for solving (\ref{cons_prob}),
\begin{subequations} \label{admm}
	\begin{align}
	\phi_{k,t}^{(l)} & =  
	\frac{ \omega_{k,t} - 2 \bm{\lambda}_{k}^{(l-1)} + \rho \sum_{j \in \mathcal{N}_k} (\phi_{k,t}^{(l-1)} + \phi_{j,t}^{(l-1)})} 
		{1 + 2 \rho |\mathcal{N}_k|}, \label{admm_phi} \\
	\bm{\lambda}_{k}^{(l)} & = \bm{\lambda}_{k}^{(l-1)} + \rho/2 \sum_{j \in \mathcal{N}_k} (\phi_{k,t}^{(l)} - \phi_{j,t}^{(l)}), \label{admm_lambda}
	\end{align}
\end{subequations}
where $l>0$ is an iteration step, $|\mathcal{N}_k|$ is the number of the neighboring nodes of node $k$,
and $\bm{\lambda}_k^{(l)} := \sum_{j \in \mathcal{N}_k} \bm{\lambda}_{kj1}^{(l)}, \forall k \in \mathcal{V}$, 
are the scaled local aggregate Lagrange multipliers. As proved in \cite{shi2014linear},
for any $\phi_{k,t}^{(0)} \in \mathbb{R}^d \times \mathbb{R}^{d \times d} \times \mathbb{R}^{d \times d}$ and $\bm{\lambda}_k^{(0)} = \bm{0}$, 
the iterations (\ref{admm_phi}) and (\ref{admm_lambda})
yield that $\phi_{k,t}^{(l)} \to \bar{\phi}_{t}$.

Note that the computation of (\ref{admm}) at node $k$ only relies on the local information $\omega_{k,t}$ and
the quantities $\{\omega_{j,t}\}_{j \in \mathcal{N}_k}$ from its neighboring nodes.
Alternating between (\ref{admm_phi}) and (\ref{admm_lambda}) for all node $k \in \mathcal{V}$,
every node can obtain the average $\bar{\phi}_{t}$ in a distributed manner.
To save the communication costs, the iterations (\ref{admm_phi})-(\ref{admm_lambda}) could be early stopped before it converges,
and a relatively small number of the maximum iteration $L$ can be set.
After the consensus step, each node $k$ can use the quantity $\phi_{k,t}^{(L)}$ to 
update the variational distributions of the global variables $x_{k,t}, X_{k,t}$ and $R_{k}$.

\subsubsection{VBM step}

Let us decompose the quantity $\phi_{k,t}^{(L)}$ as three parts: $ [(\phi_{k,t}^1)^T, (\phi_{k,t}^2)^T, (\phi_{k,t}^3)^T]^T =  \phi_{k,t}^{(L)}$.
These three parts $\phi_{k,t}^1,\phi_{k,t}^2,\phi_{k,t}^3$ are consensus results corresponding to the true sufficient statistics $\omega_{k,t}^1$, $\omega_{k,t}^2$, $\omega_{k,t}^3$, respectively.
Once obtaining the quantities  $\phi_{k,t}^1,\phi_{k,t}^2,\phi_{k,t}^3$, each node can update the posteriors $q^*(x_{k,t},X_{k,t})$ and $q^*(R_{k})$.

\paragraph{Posterior of the extended object state}
Using the quantities $\phi_{k,t}^1$ and $\phi_{k,t}^2$, each node $k$ can update its variational distribution $q^*(x_{k,t},X_{k,t})$ as
\begin{equation} \label{qxXk}
\begin{split}
& q^*(x_{k,t}, X_{k,t}) = q^*(x_{k,t}|X_{k,t})q^*(X_{k,t}) \\
& = \mathcal{N}(x_{k,t}; \hat{m}_{k,t}, \hat{P}_{k,t} \otimes X_{k,t}) \mathcal{IW}(X_{k,t}; \hat{\nu}_{k,t}, \hat{V}_{k,t}),
\end{split}
\end{equation}
with the parameters given by
\begin{subequations} \label{qxXkparams}
	\begin{align}
	\hat{m}_{k,t} & =   m_{k,t|t-1} +   ( w_t \otimes \mathbf{I}_d )e_{k,t}, \label{qxm} \\
	\hat{P}_{k,t}  & = P_{k,t|t-1} -  b_t w_t  w_t^T,  \label{qxP} \\
	\hat{\nu}_{k,t} & = \nu_{k,t|t-1} + N_t,  \label{qXnu} \\
	\hat{V}_{k,t}  & =  V_{k,t|t-1} + \frac{ N_t }{s} (\phi_{k,t}^2 - \phi_{k,t}^1 (\phi_{k,t}^1)^T) + b_t^{-1}e_{k,t} e_{k,t}^T, \label{qXV}
	\end{align}
\end{subequations}
where $ e_{k,t} \triangleq \phi_{k,t}^1 -  (H_t \otimes \mathbf{I}_d) m_{k,t|t-1}$.
\paragraph{Posterior of the measurement noise covariance} 
Using the quantity $\phi_{k,t}^3$, each node $k$ updates the variational distribution $q^*(R_k)$ as
\begin{equation}\label{qRk}
q^*(R_k) =  \mathcal{IW}(R_k; \hat{\upsilon}_{k,t}, \hat{U}_{k,t}),
\end{equation}
with the parameters given by
\begin{subequations}
	\begin{align}
	\hat{\upsilon}_{k,t} & =  \upsilon_{k,t-1} + N_t, \label{qRupsilon}\\
	\hat{U}_{k,t} & =  U_{k,t-1} + N_t \phi_{k,t}^3.  \label{qRU}
	\end{align}
\end{subequations}

For clarity,  the distributed variational Bayesian measurement update is presented in Algorithm \ref{alg_dVB},
and the distributed VB algorithm for extended object tracking (dVBEOT) is summarized in Algorithm \ref{alg_dEOT}.

\begin{algorithm}[t]
	\caption{  Distributed VB Measurement Update \label{alg_dVB}}
	\begin{algorithmic}[0]		
	\Require{ $\bar{y}_{k,t} = \frac{1}{n_k}\sum_{i=1}^{n_{k,t}}  y_{k,t}^{i}$, $m_{k,t}^{(0)} = (H_t^T \otimes \mathbf{I}_d) \bar{y}_{k,t}$, ${P}_{k,t}^{(0)} = P_{k,t|t-1}$,
			${V}_{k,t}^{(0)} = 0.1\mathbf{I}_d$, ${\nu}_t^{(0)} =d+1+0.1$, 
			${U}_{k,t}^{(0)} = U_{k,t-1} $, ${\upsilon}_{t}^{(0)} = \upsilon_{k,t-1}$.}
		\Statex
		\For{$ n \gets 0,1,2,\dots$}  
			\For{$k \in \mathcal{V}$}  \Comment{VBE step}
				\State Update $q^{(n)}(z_{k,t}^i) = \mathcal{N}(\mu_{k,t}^{i,(n)}, \Sigma_{k,t}^{i,(n)}), \forall i$ via (\ref{qzk}).
				\State Compute $\{\omega_{k,t}^{(n)}\}$ via  (\ref{omegakt}).
			\EndFor			
			\For{$ l \gets 1,2,\dots, L$} \Comment{Consensus step}
				\State  Compute $\phi_{k,t}^{(l)}$ via (\ref{admm_phi}), $\forall k\in\mathcal{V}$.
				\State  Broadcast $\phi_{k,t}^{(l)}$ to its neighbors in $\mathcal{N}_k$,  $\forall k\in\mathcal{V}$.
				\State  Compute $\bm{\lambda}_{k}^{(l)}$ via (\ref{admm_lambda}), $\forall k\in\mathcal{V}$.
			\EndFor
			\State Set $ [(\phi_{k,t}^1)^T, (\phi_{k,t}^2)^T, (\phi_{k,t}^3)^T]^T = \phi_{k,t}^{(L)},  \forall k \in \mathcal{V}$.
			\For{$k \in \mathcal{V}$}  \Comment{VBM step}
				\State  Update $q^{(n)}(x_{k,t}, X_{k,t})= \mathcal{N}(x_{k,t}; m^{(n)}_{k,t}, P_{k,t}^{(n)} \otimes $
				\State \qquad $X_{k,t}) \times \mathcal{IW}(X_{k,t}; \nu^{(n)}_{k,t}, V_{k,t}^{(n)})$ via (\ref{qxXk}).
				\State Update  $q^{(n)}(R_{k}) = \mathcal{IW}(R_k; {\upsilon}_{k,t}^{(n)}, U_{k,t}^{(n)})$ via (\ref{qRk}).
			\EndFor
		\EndFor		
		\State Set $m_{k,t|t} = m_{k,t}^{(n)}$, $P_{k,t|t} = P_{k,t}^{(n)}, \forall k$.
		\State Set $\nu_{k,t|t} = \nu_{k,t}^{(n)}$, $V_{k,t|t} = V_{k,t}^{(n)}, \forall k$.
		\State Set $\upsilon_{k,t|t} = \upsilon_{k,t}^{(n)}$, $U_{k,t|t} = U_{k,t}^{(n)}, \forall k$.
		\Statex
		\Ensure{ 
		\State \quad $P(x_{k,t} | X_{k,t}, \mathcal{Y}_{t})= \mathcal{N}(x_{k,t}; m_{k,t|t}, P_{k,t|t} \otimes X_{k,t})$. 
		\State \quad $P(X_{k,t} | \mathcal{Y}_{t}) = \mathcal{IW}(X_{k,t};\nu_{k,t|t},V_{k,t|t})$.
		\State \quad $P(R_k|\mathcal{Y}_{t}) = \mathcal{IW}(R_k; \upsilon_{k,t|t}, U_{k,t|t})$.}
	\end{algorithmic}
\end{algorithm}

\begin{algorithm}[t]
	\caption{ Distributed Variational Bayesian Algorithm for Extended Object Tracking (dVBEOT) \label{alg_dEOT}}
	\begin{algorithmic}[0]		
		
		\Require{$P_{k,1|0}=\mathbf{I}_d$, $V_{k,1|0} = 0.1\mathbf{I}_d$, $\nu_{k,1|0} = d+1+0.1$,
			$U_{k,0} = 10^{-4}\mathbf{I}_d$, $\upsilon_{k,0} = d + 1$, $\forall k \in \mathcal{N}$.}
		\Statex 
		\For{$t \gets 1,2,\dots$} \Comment{time scan}
			\State \emph{Distributed VB measurement update}:
			\State Compute $P(x_{k,t} | X_{k,t}, \mathcal{Y}_{t})$, $P(X_{k,t} | \mathcal{Y}_{t})$ and $P(R_k|\mathcal{Y}_{t})$, 
			\State $\forall k \in \mathcal{V}$ via Algorithm \ref{alg_dVB}.
			\State \emph{Local prediction}: 
			\For{$k \in \mathcal{V}$}
			\State Compute $P(x_{k,t+1},X_{k,t+1} | \mathcal{Y}_{t})$ via (\ref{local_pred}).
			\EndFor
		\EndFor 				
	\end{algorithmic}
\end{algorithm}

\begin{remark}[The case of known measurement noise covariance] \label{remark_true_noise}
If the true measurement noise covariance $R_{true}$ is known in advance, the proposed algorithm can be easily adapted to this case
and makes use of this prior information. To do this, 
we only need to replace the statistic $\langle R_k^{-1} \rangle = \hat{\upsilon}_{k,t} \hat{U}_{k,t}^{-1}$ by $R_{true}^{-1}$ in (\ref{qzmu}) and (\ref{qzsigma}) 
to update $q^*(z_{k,t}^{i})$, and eliminate the variational distribution $q(R_{k})$ in the VBM step.
Then, the distributed VB algorithm for extended object tracking with known measurement noise covariance is presented.
We examine this algorithm in the simulation.
\end{remark}

\begin{remark}[The case of neglecting the sensor error] \label{remark_neglect_noise}
If sensor error is neglectable compared to the object extension,
the proposed algorithm can also be modified by setting $R = \mathbf{0}_d$.
In detail, we  eliminate the variational distribution $q(R_{k})$ and 
replace $\langle R_k^{-1} \rangle = \hat{\upsilon}_{k,t} \hat{U}_{k,t}^{-1}$ by an extreme large value $\infty \mathbf{I}_d$.
Then the updates in (\ref{qzmu}) and (\ref{qzsigma}) reduce to $\hat{\mu}_{k,t}^i = y_{k,t}^i$ and $\hat{\Sigma}_{k,t}^i = \mathbf{0}_d$,
i.e., the latent variable $z_{k,t}^i$ is equal to the measurement $y_{k,t}^i$. 
Therefore, latent variables are unchanged with the VB iteration, 
which means we only need to perform the VB step once. After this modification, 
we found that the VB measurement update becomes the same as the measurement update in Koch's approach \cite{koch2008Bayesian}.
Therefore, as a special case of the proposed algorithm, the modified algorithm can be treated as a distributed implementation of the Koch's approach. 
\end{remark}

\begin{remark}[Group target tracking]
	As we mentioned in Introduction, a group of closely spaced targets can be considered as an extended object.
	The proposed distributed algorithm can also been used for group target tracking.
\end{remark}

\section{Simulations} \label{sec_simulation}

In this section, the performance of the proposed distributed extended object tracking 
is evaluated via numerical simulations. 

\begin{figure}[t]
 	\centering
 	\includegraphics[width=2.8in]{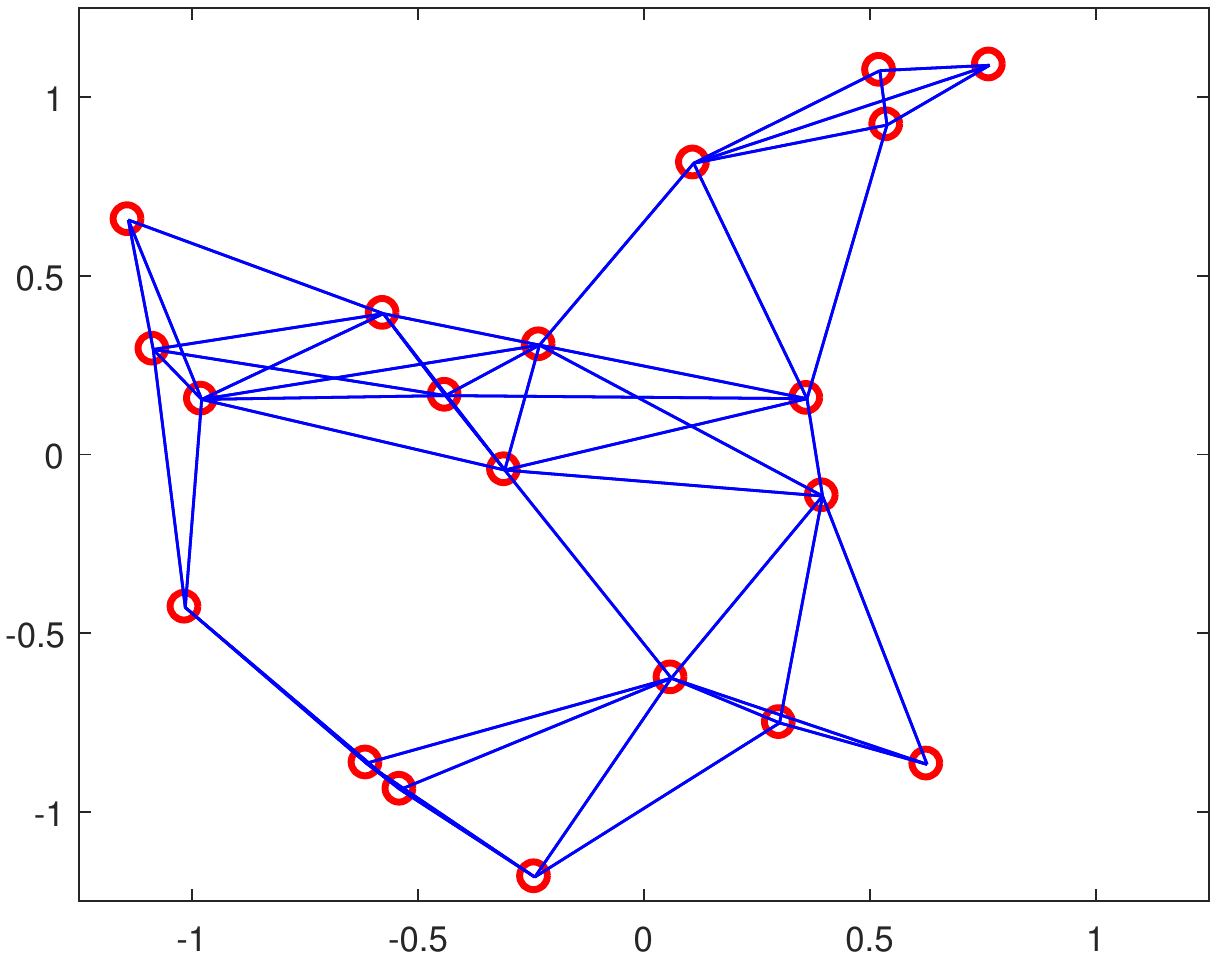}
 	\caption{Network connection.}
 	\label{fig_network}
 \end{figure}

Our numerical simulations consider a randomly generated
sensor network with $N = 20$ nodes. The nodes are randomly
placed in a $2.5 \times 2.5$ square, and the communication distance is
taken as $0.8$, as shown in \figurename{\ref{fig_network}}. 
Two scenarios were simulated: S1 for extended object tracking (EOT), and S2 for group target tracking (GTT).

For comparison, we also simulate the centralized VB algorithm for EOT (cVBEOT),
in which all data is available in a fusion center as presented in Algorithm \ref{alg_cEOT};
and the corresponding non-cooperative algorithm (non-coopVBEOT), in which each node $k$ does not perform a consensus step
and estimate the extended object state independently using local quantity $\omega_{k,t}$.
We also simulate the distributed VB for EOT with true measurement 
noise covariance (dVBEOT-with-$R_{true}$) as presented in Remark \ref{remark_true_noise};
and the algorithm neglecting the sensor error (dVBEOT-without-$R$) as presented in Remark \ref{remark_neglect_noise}.
Besides, the Koch's approach \cite{koch2008Bayesian}, which can be treated as a non-cooperative VBEOT without considering the sensor error, is also simulated.

 \subsection{Extended Object Tracking Scenario}
Extended object tracking have been applied in many different scenarios and have been evaluated using data from many different sensors
such as LIDAR, camera, radar, RGB-depth sensors, and unattended ground sensors \cite{granstrom2016extended}.
For example, in harbours, there are many vessels share the water, from small boats to large ships. To keep track of where the vessels are,
marine X-band radar can be used.

\begin{figure}[t]
	\centering
	\includegraphics[width=3.2in]{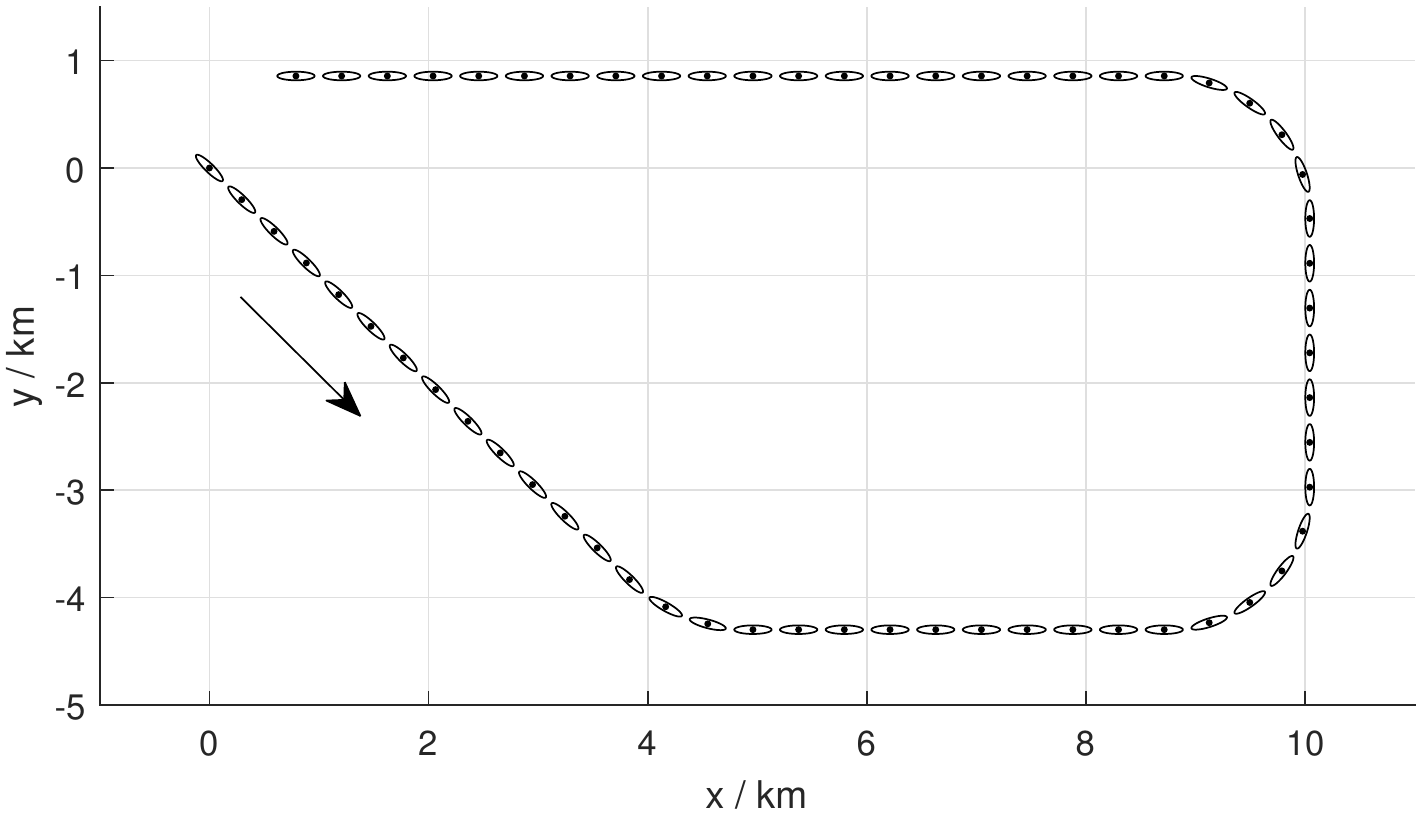}
	\caption{Trajectory of an extended object in S1.}
	\label{fig_trueEOT}
\end{figure}

In S1, we simulate the scenario of \cite{feldmann2011tracking} for maneuvering extended object tracking.
The extended object (EO) is an ellipse with diameters of $340$m and $80$m (about the size of
an aircraft carrier of the Nimitz-class \cite{feldmann2011tracking}) in the $(x,y)$-plane (thus, $d=2$).
The trajectory is shown in \figurename{\ref{fig_trueEOT}}, where
the speed was assumed constant at $27$ knots (about 50km/h), and the formation went through a $45^{\circ}$ and two $90^{\circ}$ turns.
We assumed that scattering centers are uniformly distributed over the extension $X_t$ while measurements 
are subject to a zero-mean Gaussian noise with variance $R$.
The true measurement noise is  $R_{true} = \mbox{diag}([\sigma_x^2, \sigma_y^2])$, where $\sigma_x=50$m, $\sigma_y = 50$m.
The number of measurements of each node at each time scan is Poisson distributed with mean $20$. 
The scan time of each senor is $\Delta_t = 10$s. We set the scaling factor $s=1/4$ for this scenario.

The parameters of the kinematical evolution model (\ref{kinmevolu}) are chosen as $\Sigma=1$g, $\theta = 40$s (according to \cite{koch2008Bayesian}). 
The temporal decay constant in (\ref{nu_update})
 is chosen as $\tau = \Delta_t$. The penalty parameter of the ADMM is set as $\rho = 0.5$.
 Assuming that no further information about the initial state and extension is given, we use 
 the mean of the first measurements of each node $k$ to initialize the position state and use 
 $\mathbb{E}[X_{k,1|0}]=\mathbf{I}_d$ ( with $V_{k,1|0} = 0.1\mathbf{I}_d$, $\nu_{k,1|0} = d+1+0.1$) 
 representing a circle with a radius of $1$km to initialize the extension. 

For performance evaluation of extended object estimates with ellipsoidal extents,
we use the Gaussian Wasserstein Distance (GWD) metric as the measure of performance, which is best choice for comparing elliptic shapes \cite{yang2016metrics}.
The squared $L_2$ Wasserstein distance between two multivariate Gaussian is defined as \cite{givens1984class}
\begin{equation}
\begin{split}
d_{GW}(\mathcal{N}_{y},\mathcal{N}_{\hat{y}})^2 =& ||\mu_{y} - \mu_{\hat{y}}||^2 \\
& + \operatorname{tr}\left(\Sigma_{y} + \Sigma_{\hat{y}} - 2 (\Sigma_{y} ^{\frac{1}{2}} \Sigma_{\hat{y}}  \Sigma_{{y}} ^{\frac{1}{2}})^{\frac{1}{2}}\right),
\end{split}
\end{equation}
where $\mathcal{N}_{y}(\mu_{y}, \Sigma_{y})$ is the groundtruth ellipse and $\mathcal{N}_{\hat{y}}(\mu_{\hat{y}},  \Sigma_{\hat{y}})$ is the estimated ellipse.
The comparison results are the root Gaussian Wasserstein error (RGWE) over $N_s$ independent Monte Carlo runs with randomly generate samples,
calculated as follows:
\begin{equation}
\mbox{RGWE}_{k,t} = 
\left( \frac{1}{N_s} \sum_{l=1}^{N_s} d_{GW}(\mathcal{N}_{k,t}, \mathcal{N}_{t})^2 \right)^{\frac{1}{2}},
\end{equation}
where $\mathcal{N}_{t}((H_t \otimes \mathbf{I}_d) x_{t}, s X_t)$ is the groundtruth ellipse and 
$\mathcal{N}_{k,t}((H_t \otimes \mathbf{I}_d) \hat{x}_{k,t}, s \hat{X}_{k,t})$ is the estimated ellipse at time scan $t$ at node $k$.
For the distributed algorithms, the final results are averaged over all nodes.

\begin{figure}[t]
	\centering
	\includegraphics[width=3in]{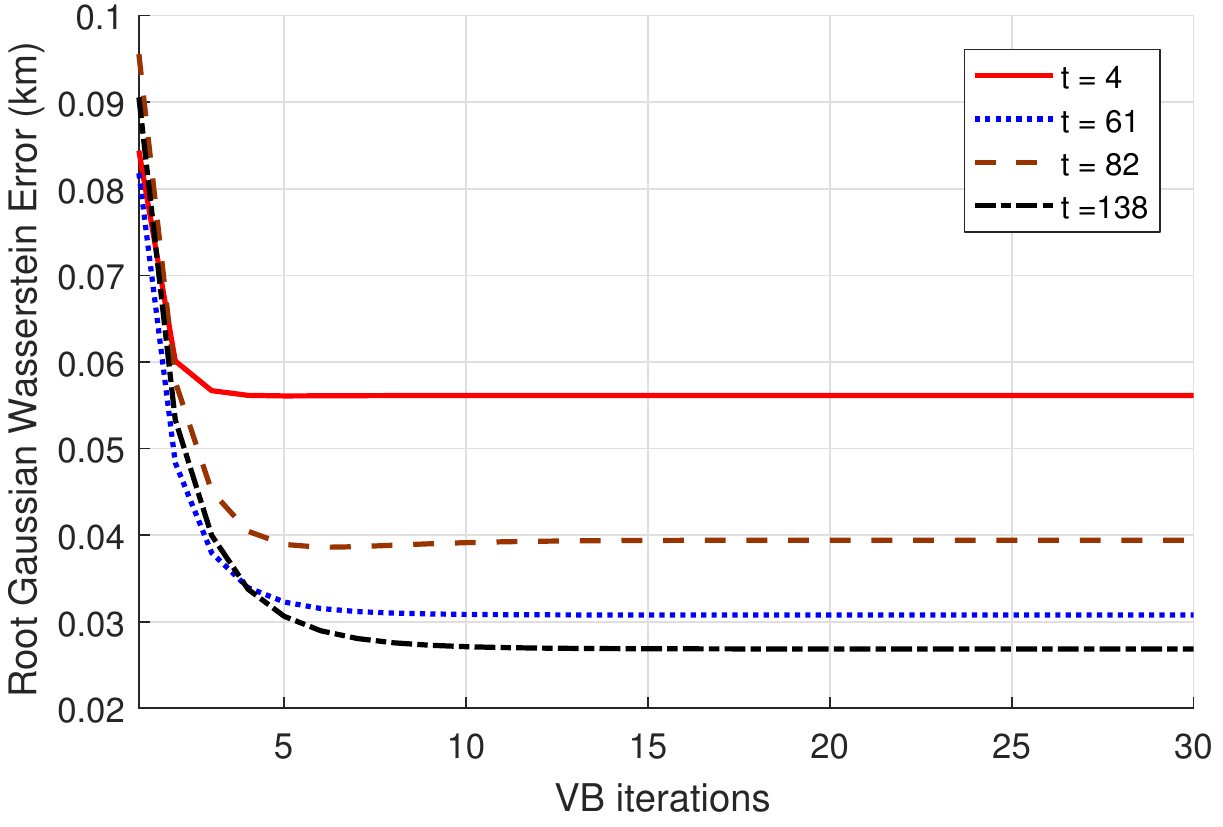}
	\caption{The RGWE evolution of the dVBEOT at a randomly selected node ($k=2$) at four time scans ($t=4, 61, 82, 138$). We set $L=50$.}
	\label{fig_vb_converge}
\end{figure}

\begin{figure}[t]
	\centering
	\includegraphics[width=3in]{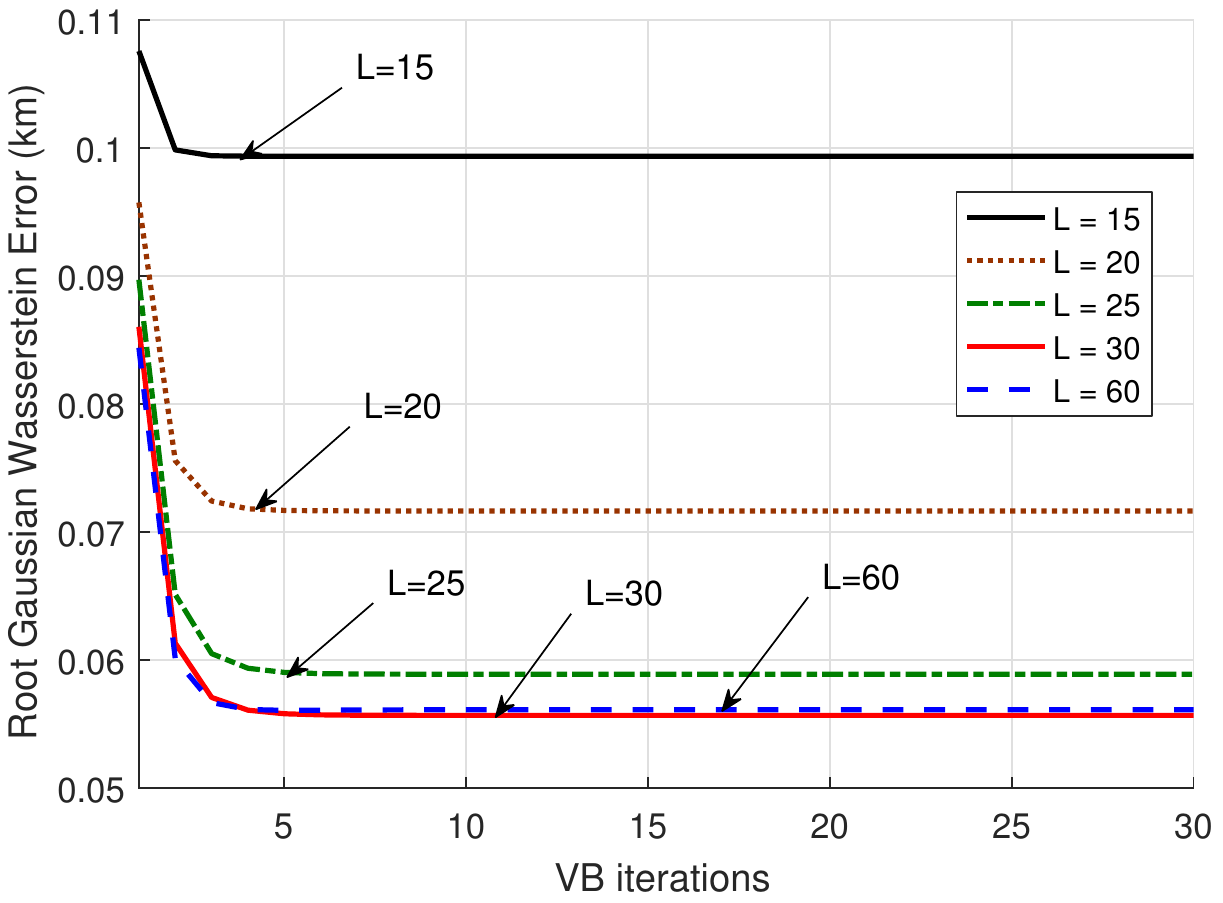}
	\caption{The RGWE evolution of the dVBEOT with different number of inner iterations ($L=15, 20, 25, 30, 60$) at a randomly selected node ($k=2$) at time scan $t=4$.}
	\label{fig_admm_L}
\end{figure}

\subsubsection{Convergence Study}
We check the convergence of the proposed dVBEOT algorithm. 
We first evaluated  the  root Gaussian Wasserstein error (RGWE) of the dVBEOT with the VB iterations at different time scans, 
and four of them are shown in \figurename{\ref{fig_vb_converge}}. The number of the ADMM iterations in the consensus step is set as $L = 50$. 
We observed that the proposed VB algorithm converges after only $10$ iterations at all time scans.
We then evaluated the RGWE performance with different number of the inner iterations $L$.
As shown in \figurename{\ref{fig_admm_L}}, the RGWE is getting smaller and smaller with the increasing 
of the number of the ADMM iterations $L$. When $L > 30$, the RGWE is almost unchanged, which means 
all nodes have achieved consensus on  the expected  sufficient statistics $\{\phi_{k,t}\}$ in every VB step. 
In the following simulations, we set $L=30$ and run the VB step $20$ times.

\subsubsection{Tracking Performance}
\begin{figure*}[t]
	\centering
	\includegraphics[width=6.8in]{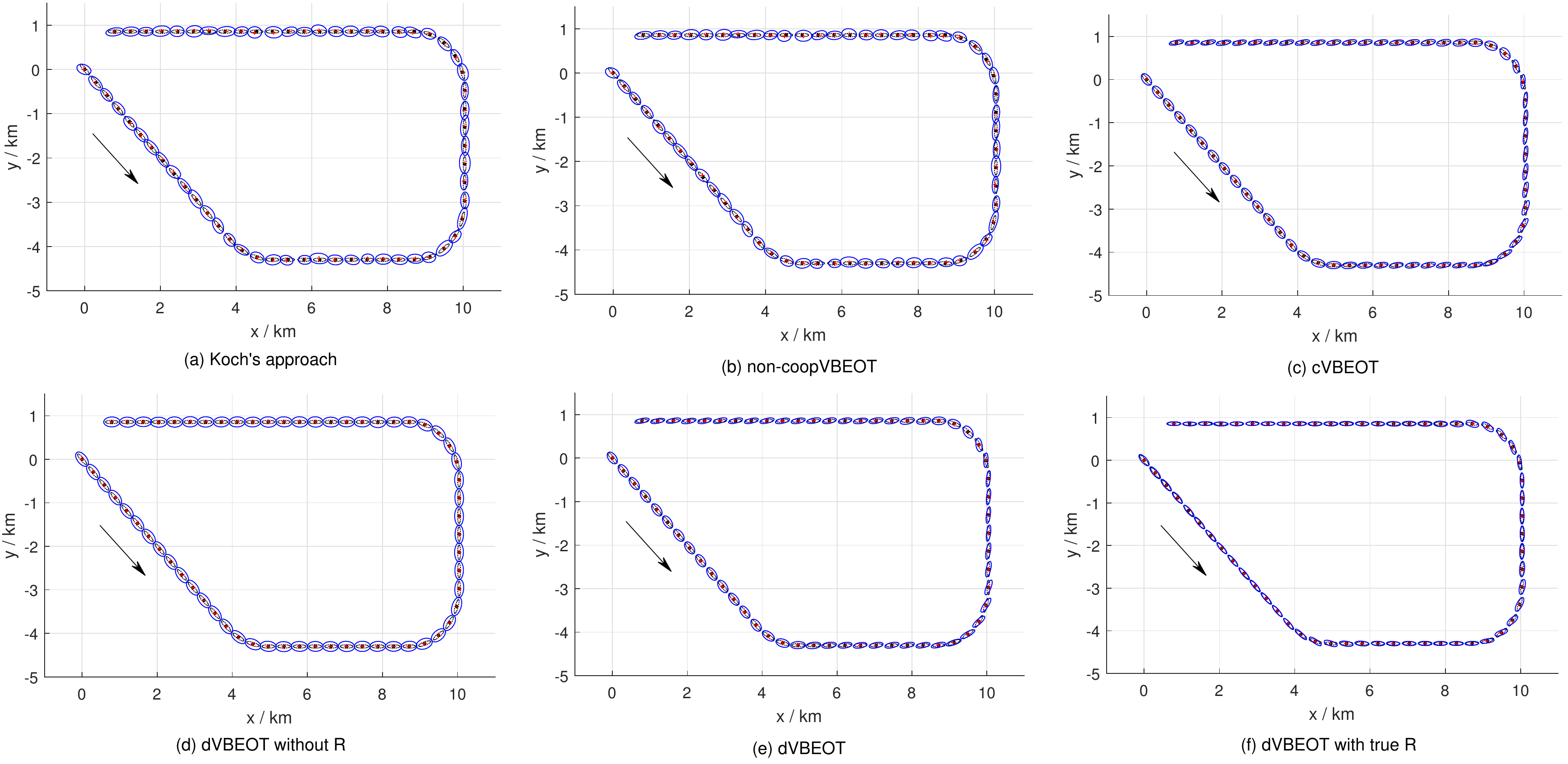}
	\caption{Tracking results of all six algorithms for the sensor data of \figurename{\ref{fig_trueEOT}} at a randomly selected node ($k=2$). 
		The estimated extension is represented by $90\%$-confidence ellipse 
		Shown are, for each time scan $t$, the true centroid (block point), the groundtruth ellipse (black dotted line), the estimated centroid(red $+$),
		and the estimated extension (blue thick line). }
	\label{fig_all_ellipse_EOT}
\end{figure*}

\begin{figure}[t]
	\centering
	\includegraphics[width=2.8in]{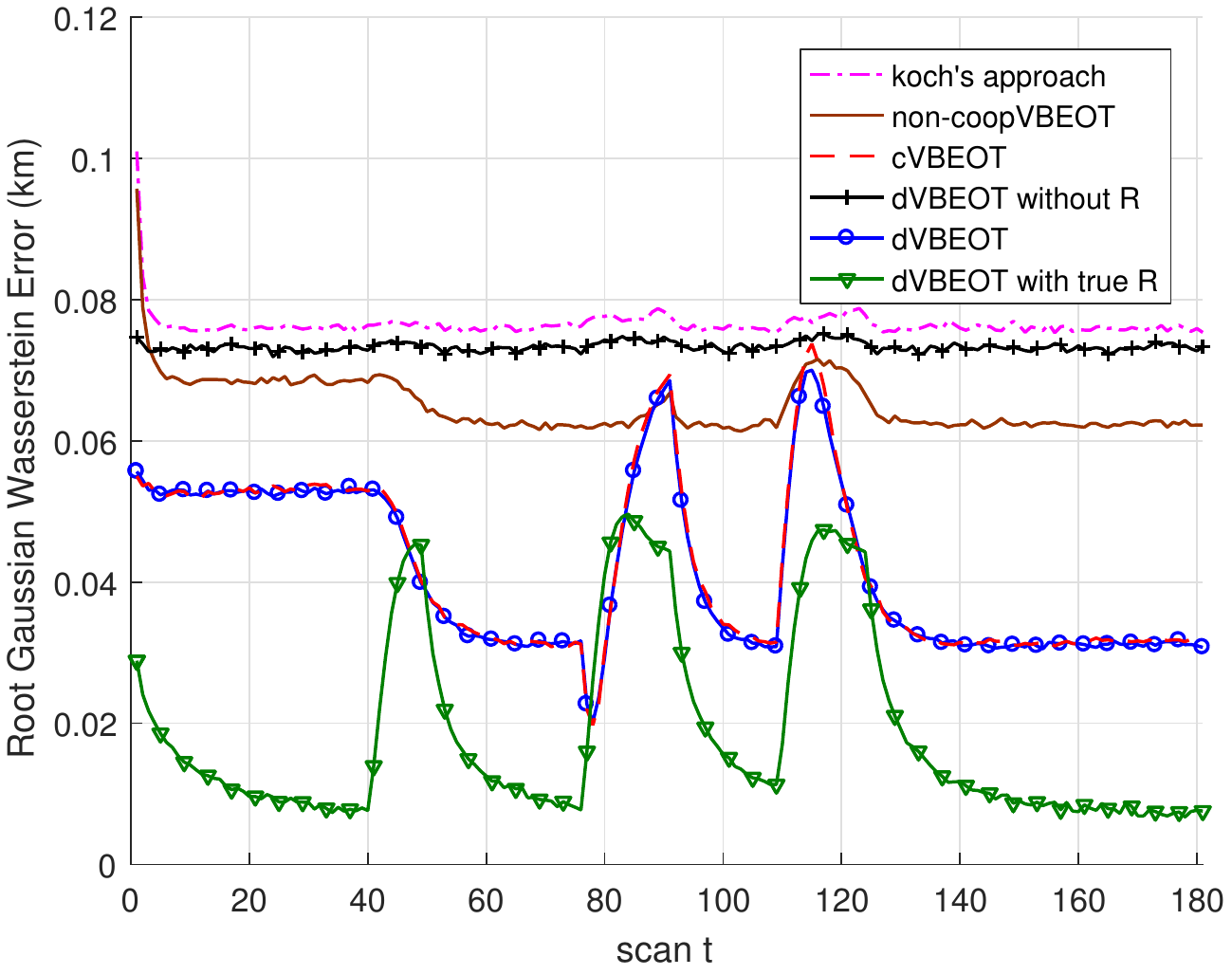}
	\caption{The root Gaussian Wasserstein error (RGWE) of all six algorithms for extended object tracking in S1.}
	\label{fig_all_rgwe_EOT}
\end{figure}

We compared the proposed dVBEOT algorithm with Koch's approach, the non-coopVBEOT,
the cVBEOT, the dVBEOT-without-$R$  and the dVBEOT-with-$R_{true}$.
\figurename{\ref{fig_all_ellipse_EOT}} shows the tracking results of all six algorithms at a 
randomly selected node $k$. The estimated extension is represented by $90\%$-confidence ellipse.
In \figurename{\ref{fig_all_rgwe_EOT}}, we summarized the RGWE results of each algorithm over $N_s = 100$ Monte Carlo runs.
As shown in \figurename{\ref{fig_all_ellipse_EOT}--\ref{fig_all_rgwe_EOT}}, 
both Koch's approach and the dVBEOT-without-$R$ overestimated the extension and have highest estimation error, 
since they do not consider the actual measurement noise. 
The non-coopVBEOT has a better RGWE performance than 
that of Koch's approach and that of the dVBEOT-without-$R$, but it is still 
not so good. The proposed dVBEOT performs much better than the above three mentioned algorithms, and 
is almost as good as the corresponding centralized algorithm (cVBEOT), which utilizes all measurements
in a fusion center. This shows that the proposed distributed algorithm can make use of the information from all nodes,
which verifies the effectiveness of the proposed algorithm. 
The dVBEOT-with-$R_{true}$, which utilizes the prior information 
about the true measurement noise, has the best tracking performance and the lowest RGW error,
and its estimated position and extension are almost same as the ground truth except the case when 
the formation of the extended object went turns (at time scan  $t=40, 80, 110$).
Nevertheless, after the formation went turns, the proposed algorithm can amend its extension very quickly
and still achieve very low RGW error.

\subsection{Group target Tracking Scenario}

 \begin{figure}[t]
 	\centering
 	\includegraphics[width=2.8in]{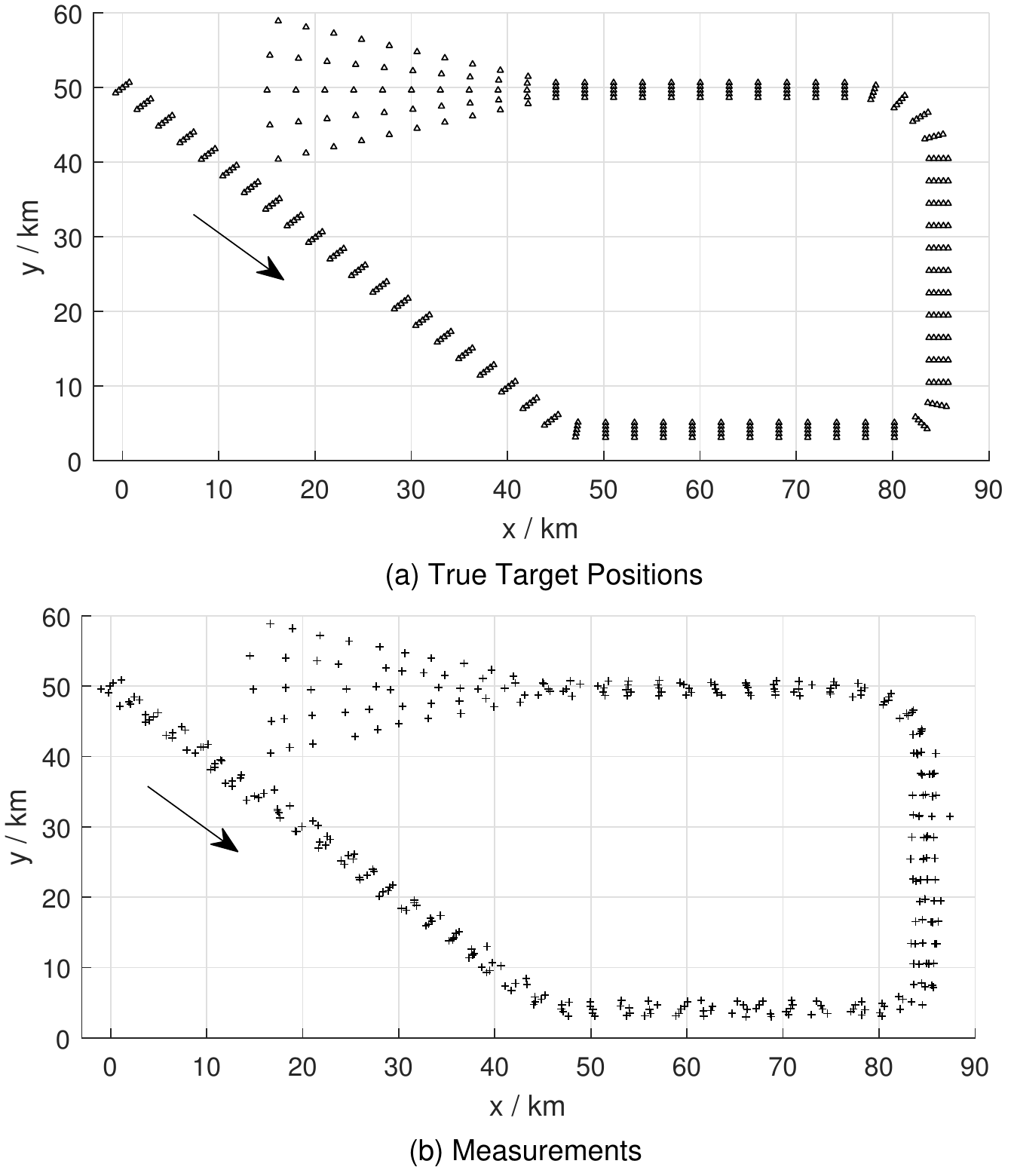}
 	\caption{Group tracking scenario with $5$ targets. (a) The true target position. (b) The 
 		measurements of a random selected node.}
 	\label{fig_truegttdata}
 \end{figure}
 
In S2, we test the performance of the proposed algorithm on the group target tracking problem.
In this scenario, a group of uniformly spaced five targets moves in a plane with the trajectories shown in \figurename{\ref{fig_truegttdata}}. 
Five individual targets fly with constant speed $v = 300$ m/s in the $(x,y)$-plane (thus, $d=2$). 
The targets were arranged in a line with $500$m distance between neighboring targets, 
where the formation first went through a $45^{\circ}$ and two $90^{\circ}$ turns (with radial accelerations
$2$g, $2$g, $1$g, respectively) before performing a split-off maneuver. The scan time of each sensor is $\Delta_t=10$s.
The true measurement noise is  $R_{true} = \mbox{diag}([\sigma_x^2, \sigma_y^2])$, where $\sigma_x=500$m, $\sigma_y = 100$m.
We assume that each node has a probability of detection $P_d = 80\%$ for each target at each time scan $t$.
We set $L=50$, run the VB step $80$ times and set other parameters the same as S1.

\begin{figure*}[t]
	\centering
	\includegraphics[width=6.8in]{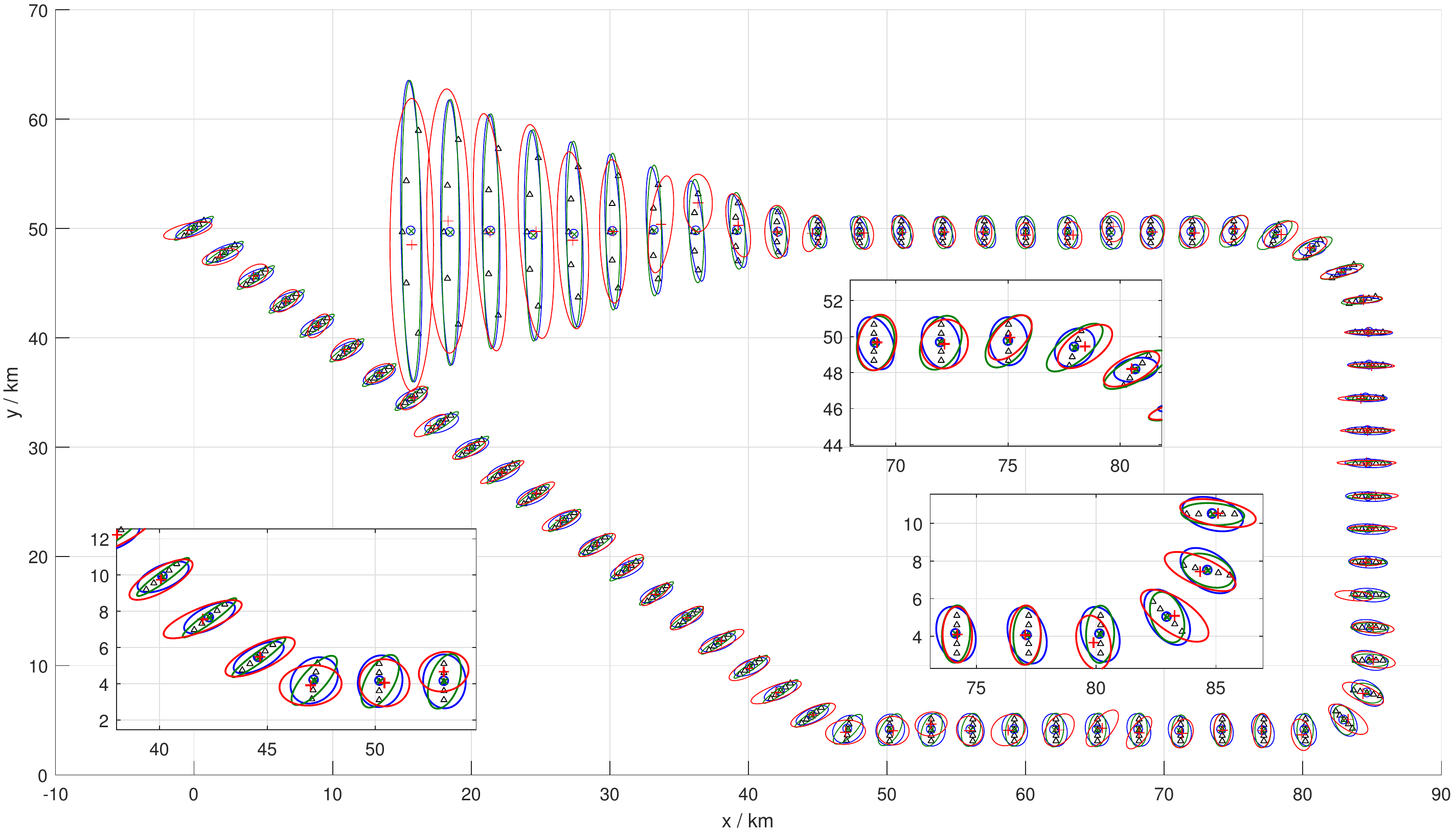}
	\caption{The tracking results for group target tracking in S2. 
		The estimated extension is represented by $90\%$-confidence ellipse.
		Shown are, the true target positions ($\triangle$),  
		the estimated centroid (red $+$) and the estimated extension (red line) of Koch's approach;
		the estimated centroid (blue $\Circle$) and the estimated extension (blue line) of the dVBEOT;
		the estimated centroid (black $+$) and the estimated extension (black line) of the dVBEOT with $R_{true}$.}
	\label{fig_gtt_all}
\end{figure*}

For ease of comparison, \figurename{\ref{fig_gtt_all}} shows the tracking results of three algorithms: Koch's approach, the 
dVBEOT and the dVBEOT-with-$R_{true}$. It is shown that the Koch's approach often has large errors on the centroid and extension 
due to its sensitive to missed detections. While, the dVBEOT has a significant improvement on extended state estimation,
especially on extension estimation. This is because the nodes in sensor networks can gather more measurements than a single node to give
a more complete description on the extended object, and the proposed distributed algorithm can obtain this information through the
cooperation among nodes. Besides, the dVBEOT can avoid overestimating the object size to some extent by estimating the sensor errors.
Moreover, it is shown that the dVBEOT-with-$R_{true}$ can effectively utilize the prior information about the true measurement noise, 
and has best extended object state estimation performance.
The comparison results demonstrate the effectiveness of the proposed distributed algorithms to the group target tracking.

\section{Conclusion} \label{sec_conclusion}
Consider the case that an object is spatially structured, 
we propose a distributed tracking algorithm for extended objects in senor networks. We formulate a distributed
Bayesian model for extended object tracking with unknown measurement noise.
Based on the variational Bayesian methods, we derive 
a new measurement update for the estimation of the kinematic state and extension
as well as the measurement noise covariance.
Then, using the ADMM technique, we derive the corresponding distributed Bayesian tracking algorithm.
Numerical simulations on both extended object tracking and group target tracking demonstrate
that the proposed dVBEOT algorithm has superior performance for cases where sensor errors cannot
be neglected any more in comparison with object extension. 
The proposed distributed algorithm performs as good as the corresponding centralized algorithm.
Moreover, when the true measurement noise covariance is available, the dVBEOT can utilize
this prior information and achieve much better performance.

\appendix

\subsection{Wishart and Inverse Wishart Distributions} \label{wishart}

1) Wishart density: $\mathcal{W}(X; n, W)$ denotes a Wishart density defined over the matrix $X \in \mathbb{S}_{++}^d$ with scalar degrees of freedom $n > d-1$
and parameter matrix $W \in \mathbb{S}_{++}^d$,
\begin{equation}
\mathcal{W}_d(X; n, W) = \frac{ |W|^{-\frac{n}{2}} |X|^{\frac{n-d-1}{2}}}{ 2^{\frac{nd}{2}}  \Gamma_d(\frac{n}{2}) } \mbox{etr}\left(-\frac{1}{2} W^{-1} X\right),
\end{equation}
where $\mbox{etr}(\cdot) = \exp(\operatorname{tr}(\cdot))$  is exponential of the matrix trace, and $\Gamma_d(\cdot)$ 
is the multivariate gamma function. The expectation of $X$ is given by $\mathbb{E}[X] = n W$.

2) Inverse Wishart density: If $X \sim \mathcal{W}_d(X; n, W)$, then the random variance matrix $\Sigma = X^{-1}$ has an inverse Wishart distribution, denoted by
\begin{equation}
\mathcal{IW}_d(\Sigma; n, W) 
= \frac{ |W|^{\frac{n}{2}} |\Sigma|^{-\frac{n+d+1}{2}} }{2^{\frac{n d}{2}} \Gamma_d(\frac{n}{2})}  \mbox{etr}\left(-\frac{1}{2} \Sigma^{-1} W\right),
\end{equation}
with  scalar degrees of freedom $n > d-1$ and parameter matrix $W \in \mathbb{S}_{++}^d$.
The expectation of $\Sigma$ is given by $\mathbb{E}[\Sigma]=W/(n-d-1)$ when $n > d + 1$.

\subsection{proof of (\ref{qxX})} \label{appendix_proof_xX}

From (\ref{qxXform}), grouping terms involving $x_t$, we have
\begin{equation} \label{qxform}
\begin{split}
 & \ln^* q(x_t |X_t) \\
 &  \propto-\frac{1}{2}x_t^T \left((P_{t|t-1} \otimes X_t)^{-1}  \right. \\
 &  \left. \qquad \qquad +   N_t   (H_t \otimes \mathbf{I}_d)^T (s X_t)^{-1} (H_t \otimes \mathbf{I}_d) \right)x_t   \\
&  \quad +  \big( m_{t|t-1}^T (P_{t|t-1} \otimes X_t)^{-1} +N_t\bar{z}_t^T(s X_t)^{-1} (H_t \otimes \mathbf{I}_d) \big) x_t  \\
 & \propto-\frac{1}{2}x_t^T \big(  (P_{t|t-1}^{-1}  + \frac{N_t}{s}H_t^T H_t)\otimes X_t^{-1} \big)x_t \\
& \quad + x_t^T \big(  (P_{t|t-1}^{-1} \otimes X_t^{-1}) m_{t|t-1} +\frac{N_t}{s} (H_t^T \otimes X_t^{-1})\bar{z}_t \big),
\end{split}
\end{equation}
where the second equation is derived by using the fact  that $h$ is a row vector and
\begin{subequations}
	\begin{align}
	& (P_{t|t-1} \otimes X_t)^{-1}  = P_{t|t-1}^{-1} \otimes X_t^{-1},  \\
	& (H_t \otimes \mathbf{I}_d)^T X_t^{-1} (H_t \otimes \mathbf{I}_d) = (H_t^T H_t) \otimes X_t^{-1}, \\
	& X_t^{-1} (H_t \otimes \mathbf{I}_d) =  H_t \otimes X_t^{-1} .
	\end{align}
\end{subequations}

From (\ref{qxform}), we conclude that $q^*(x_t |X_t)$ is Gaussian distributed,
\begin{equation}
q^*(x_t |X_t) = \mathcal{N}(x_t; \hat{m}_t, \hat{P}_t \otimes X_t),
\end{equation}
with the parameters
\begin{subequations}
	\begin{align}
	\hat{P}_t  & =   (P_{t|t-1}^{-1}  + \frac{N_t}{s}H_t^T H_t)^{-1}, \\
	\hat{m}_t & =  (\hat{P}_t \otimes X_t) 
	( ( P_{t|t-1}^{-1} \otimes X_t^{-1}) m_{t|t-1} + \frac{N_t}{s} (H_t^T \otimes X_t^{-1})\bar{z}_t ) \nonumber\\
	& = ((\hat{P}_t P_{t|t-1}^{-1}) \otimes \mathbf{I}_d ) m_{t|t-1} + 
	( ( \frac{N_t}{s}  \hat{P}_t H_t^T) \otimes \mathbf{I}_d ) \bar{z}_t.
	\end{align}
\end{subequations}

Furthermore, using the matrix inversion lemma, we can further simply the updates,
\begin{equation}
\begin{split}
&  \frac{N_t}{s}  \hat{P}_t H_t^T =  \frac{N_t}{s}  (P_{t|t-1}^{-1}  + \frac{N_t}{s}H_t^T H_t)^{-1}H_t^T \\
& \quad = \frac{N_t}{s}   ( \mathbf{I}_3  + \frac{N_t}{s} P_{t|t-1} H_t^T H_t)^{-1} P_{t|t-1}H_t^T\\
&\quad = \frac{N_t}{s} (\mathbf{I}_3 - P_{t|t-1}H_t^T(\frac{s}{N_t} + H_t P_{t|t-1}H_t^T)^{-1} h) P_{t|t-1}H_t^T \\
&\quad = \frac{N_t}{s}  P_{t|t-1}H_t^T ( \mathbf{I}_3 - (\frac{s}{N_t} + H_t P_{t|t-1}H_t^T)^{-1} H_t P_{t|t-1}H_t^T) \\
& \quad= \frac{N_t}{s}  P_{t|t-1}H_t^T (\mathbf{I}_3 + \frac{N_t}{s} H_t P_{t|t-1}H_t^T )^{-1} \\
& \quad = P_{t|t-1}H_t^T(\frac{s}{N_t} + H_t P_{t|t-1}H_t^T)^{-1} \\
&\quad = w_t,
\end{split}
\end{equation}
and
\begin{equation}
\begin{split}
\hat{P}_t P_{t|t-1}^{-1} & = (P_{t|t-1}^{-1}  + \frac{N_t}{s}H_t^T H_t)^{-1} P_{t|t-1}^{-1} \\
& =   (\mathbf{I}_3  + \frac{N_t}{s} P_{t|t-1}H_t^T H_t)^{-1} \\
& = \mathbf{I}_3 - P_{t|t-1}H_t^T(\frac{s}{N_t} + H_t P_{t|t-1}H_t^T)^{-1} H_t\\
& = \mathbf{I}_3 - w_t H_t,      
\end{split}
\end{equation}

where we define
\begin{subequations}
	\begin{align}	
	w_t &\triangleq P_{t|t-1}H_t^T b_t^{-1},  \label{Wt}\\
	b_t &\triangleq \frac{s}{N_t} + H_t P_{t|t-1}H_t^T.
	\end{align}
\end{subequations}
Note that $w_t \in \mathbb{R}^{3 \times 1}$ is a column vector, and $b_t \in \mathbb{R}$ is a scalar.
Based on the above results, we can rewritten the parameters as
\begin{equation}
\begin{split}
\hat{m}_t & = ((\mathbf{I}_3 - w_t H_t) \otimes \mathbf{I}_d ) m_{t|t-1} +    ( w_t \otimes \mathbf{I}_d ) \bar{z}_t \\
& = ((\mathbf{I}_{3d} -( w_t \otimes \mathbf{I}_d ) (H_t  \otimes \mathbf{I}_d )) m_{t|t-1} +    ( w_t \otimes \mathbf{I}_d ) \bar{z}_t \\
& =  m_{t|t-1} +   ( w_t \otimes \mathbf{I}_d )( \bar{z}_t -  (H_t  \otimes \mathbf{I}_d ) m_{t|t-1} ),
\end{split}
\end{equation}
and 
\begin{equation}
\begin{split}
\hat{P}_t  & =    (\mathbf{I}_3 - w_t H_t)P_{t|t-1} = P_{t|t-1} -  w_t b_t w_t^T.
\end{split}
\end{equation}

Thus, the variational distribution $q^*(x_t, X_t)$ in (\ref{qxXform}) can be rewritten as 
\begin{equation} \label{qxXform2}
\begin{split}
& \ln q^*(x_t, X_t) \\
&  =  \ln  \mathcal{N}(x_t; \hat{m}_t, \hat{P}_t \otimes X_t) + \frac{1}{2} \hat{m}_t^T(\hat{P}_t \otimes X_t)^{-1} \hat{m}_t \\
 & \quad  - \frac{1}{2} \operatorname{tr}( X_t^{-1}  (\frac{N_t}{s} \mathbf{S}_t + V_{t|t-1})  ) - \frac{1}{2} \frac{N_t}{s} \bar{z}_t^T X_t^{-1} \bar{z}_t\\
& \quad  -\frac{1}{2}   m_{t|t-1}^T (P_{t|t-1}^{-1} \otimes X_t^{-1})m_{t|t-1}   \\
& \quad  -\frac{(N_t + \nu_{t|t-1})+d+1}{2} \ln |X_t| + c.
\end{split}
\end{equation}
Let us define 
\begin{equation}
\begin{split}
\Delta_t & \triangleq \frac{M}{s} \bar{z}_t^T X_t^{-1} \bar{z}_t  + m_{t|t-1}^T (P_{t|t-1}^{-1} \otimes X_t^{-1})m_{t|t-1} \\
&  \quad  - \hat{m}_t^T(\hat{P}_t \otimes X_t)^{-1} \hat{m}_t.
\end{split}
\end{equation}

From (\ref{qxform}), we have
\begin{equation}
\begin{split}
	& \hat{m}_t^T(\hat{P}_t \otimes X_t)^{-1} \hat{m}_t \\
	& = (m_{t|t-1} +   ( w_t \otimes \mathbf{I}_d )( \bar{z}_t -  (H_t  \otimes \mathbf{I}_d ) m_{t|t-1} ))^T \\
	&  \quad \times ( (P_{t|t-1}^{-1} \otimes X_t^{-1}) m_{t|t-1} + \frac{M}{s} (H_t^T \otimes X_t^{-1})\bar{z}_t).
\end{split}
\end{equation}
Moreover, 
\begin{subequations}
	\begin{align}
	 ( w_t^T \otimes \mathbf{I}_d ) (P_{t|t-1}^{-1} \otimes X_t^{-1})  & =  b_t^{-1} H_t \otimes X_t^{-1}, \\	 
	 ( w_t^T \otimes \mathbf{I}_d ) (H_t^T \otimes X_t^{-1})  & =  w_t^T H_t^T  X_t^{-1},  \\
	 H_t^T \otimes X_t^{-1} & = (H_t^T \otimes \mathbf{I}_d) X_t^{-1}.
	\end{align}
\end{subequations}

Therefore, we have
\begin{equation}
\begin{split}
\Delta_t & =   \frac{M}{s} \bar{z}_t^T X_t^{-1} \bar{z}_t -  \frac{M}{s}  \bar{z}_t^T  X_t^{-1}  (H_t \otimes \mathbf{I}_d) m_{t|t-1} \\
& \quad - ( \bar{z}_t - (H_t   \otimes \mathbf{I}_dm_{t|t-1} ) )^T  ( b_t^{-1} X_t^{-1} (H_t \otimes \mathbf{I}_d) m_{t|t-1} \\
&  \qquad   + \frac{M}{s}  w_t^T H_t^T  X_t^{-1} \bar{z}_t) \\
& =   \bar{z}_t^T \frac{M}{s}(  1 -   w_t^T H_t^T   )X_t^{-1}\bar{z}_t \\
& \quad + \bar{z}_t^T(- \frac{M}{s} -b_t^{-1} + \frac{M}{s}  H_t w_t  )X_t^{-1} (H_t \otimes \mathbf{I}_d) m_{t|t-1} \\
& \quad + m_{t|t-1}^T(H_t^T \otimes \mathbf{I}_d)  b_t^{-1}X_t^{-1} (H_t \otimes \mathbf{I}_d) m_{t|t-1}.
\end{split}
\end{equation}
Using (\ref{Wt}), we have $\frac{M}{s}(1 -   w_t^T H_t^T)  = b_t^{-1}$ and $- \frac{M}{s} -b_t^{-1} + \frac{M}{s}  H_t w_t = -2b_t^{-1}$.
Thus,
\begin{equation}
\Delta_t = (\bar{z}_t - (H_t \otimes \mathbf{I}_d) m_{t|t-1})^T(b_tX_t)^{-1}(\bar{z}_t - (H_t \otimes \mathbf{I}_d) m_{t|t-1}).
\end{equation}
Let us define
\begin{equation}
\mathbf{K}_t \triangleq b_t^{-1}(\bar{z}_t - (H_t \otimes \mathbf{I}_d) m_{t|t-1})(\bar{z}_t - (H_t \otimes \mathbf{I}_d) m_{t|t-1})^T.
\end{equation}
We can rewrite  $q^*(x_t, X_t)$ in (\ref{qxXform2}) as 
\begin{equation} \label{qxXform3}
\begin{split}
& \ln q^*(x_t, X_t) \\
&  =  \ln  \mathcal{N}(x_t; \hat{m}_t, \hat{P}_t \otimes X_t)  \\
& \quad  - \frac{1}{2} \operatorname{tr}( X_t^{-1}  (\frac{N_t}{s} \mathbf{S}_t + V_{t|t-1} + \mathbf{K}_t)  )  \\
& \quad  -\frac{(N_t + \nu_{t|t-1})+d+1}{2} \ln |X_t| + c \\
& = \ln  \mathcal{N}(x_t; \hat{m}_t, \hat{P}_t \otimes X_t)  + \ln \mathcal{IW}(X_t; \hat{\nu}_t, \hat{V}_t),
\end{split}
\end{equation}
where 
\begin{subequations}
	\begin{align}
	\hat{\nu}_t & = N_t + \nu_{t|t-1}\\
	\hat{V}_t  & =  \frac{M}{s} \mathbf{S}_t + \mathbf{K}_t + V_{t|t-1}.
	\end{align}
\end{subequations}
Thus, we obtain (\ref{qxX}).



\end{document}